\documentclass[11pt,english]{article}

\usepackage[latin9]{inputenc}
\usepackage{geometry}
\usepackage[greek,USenglish]{babel}
\usepackage{verbatim}
\usepackage{prettyref}
\usepackage{amsmath}
\usepackage{amssymb}
\usepackage{graphicx}
\usepackage{setspace}
\usepackage{esint}
\usepackage{marvosym}
\usepackage{color}
\definecolor{darkgreen}{rgb}{0,0.5,0}
\definecolor{darkblue}{rgb}{0,0,0.6}
\definecolor{purple}{rgb}{0.4,.2,0.7}
\usepackage[colorlinks=true,citecolor=darkgreen,linkcolor=darkblue,urlcolor=purple]{hyperref}

\numberwithin{figure}{section}
\numberwithin{equation}{section}
\numberwithin{table}{section}

\begin{document}

\vspace*{1.5cm}
\begin{center}
{ \LARGE {\textsc{De Sitter Musings}}}
\vspace*{1.5cm}
 
Dionysios Anninos$^{\dag}$
\vspace*{1cm}

{\it $^\dag$ Stanford Institute of Theoretical Physics, Stanford University, Stanford, CA 94305-4060, USA}\\
\end{center}
\vspace*{1.5cm}

\begin{abstract}
We discuss some of the issues arising when considering asymptotically de Sitter spacetimes, and attempts to address them. Our development begins at the classical level, where several initial value problems are discussed, and ends with several proposals for holography in asymptotically de Sitter spacetimes. Throughout the paper we give a review of some basic notions such as the geometry of the Schwarzschild-de Sitter black hole, the Nariai limit, and quantum field theory in a fixed de Sitter background. We also briefly discuss some semiclassical aspects such as the nucleation of giant black holes and the Hartle-Hawking wavefunctional. We end by giving an overview of some open questions. An emphasis is placed on the differences between a static patch observer confined to live in a thermal cavity and the metaobserver who has access to a finite region of the future boundary.  
\end{abstract}
\vspace*{1.5cm}

\newpage

\tableofcontents

\section{Introduction - A tale of two observers}

The universe is expanding. Moreover, it is expanding at an accelerating pace \cite{Perlmutter:1998np,Riess:1998cb}, as it had once done before in its earliest of stages. If this expansion persists, we will eventually head toward a cold and lonely world. The largest of structures, beginning with superclusters and followed by galactic clusters, will slowly dilute away from our own Local Group in the next hundred billion years or so, until we are left with only a merger galaxy of the Milky Way and Andromeda \cite{Nagamine:2002wi,Busha:2003sz}. After even longer time scales, around a trillion years \cite{Busha:2003sz}, all cosmic radiation will have stretched to sizes beyond the horizon. The stelliferous era will come to an end in about a hundred trillion years. Our whole observable world will be governed by thermal and quantum fluctuations of a structureless cosmic cavity at a Hawking temperature of $\sim 10^{-29} \; \text{K}$. The main source of energy will come from the non-diluting cosmological constant, which already constitutes $\sim 0.7$ of the energy in the present day universe. We will be separated from all other worlds by a great cosmic horizon, some $17$ billion light years in size. What exotic and mysterious physics, if any, might spawn \cite{Coleman:1980aw} from the far future of the cosmic cavity itself? As we shall see, such observations about our universe lie at the heart of a host of theoretical questions which have baffled theoretical physicists to this day. 

We have observational evidence for two periods of exponential expansion. The first is the early inflationary era during which the universe dramatically expanded in size moments after the big bang and planted the seeds for the eventual formation of the structure we observe in the night sky. A given observer in an asymptotically de Sitter universe, which due to its exponential expansion surrounds the observer by a cosmological horizon, does not ordinarily have access to the data on the spatial slice at the infinite future $\mathcal{I}^+$. Thus, the physical meaning of $\mathcal{I}^+$, and more generally the data outside the observer's horizon, is rendered suspect. But inflation came to a sudden end, followed by a period of reheating and the eventual formation of large scale structure. This allows us to access information on late time slices of this approximately de Sitter phase. This data would have become the data on $\mathcal{I}^+$, had inflation persisted forever, but is now reentering the visible universe. Thus, we are in a sense `metaobservers' of this early de Sitter era. Indeed, the scale invariant power spectrum of the CMB is direct evidence of the structure of the approximate $\mathcal{I}^+$ of inflation. But the story doesn't end there. Many models of inflationary cosmology are sourced by a scalar field which may experience quantum fluctuations back up its inflationary potential. Hence, the global structure of the reheating surface may be far wilder than a smooth spacelike slice which we can access, and may lead to a never-ending production of post-reheating observers like ourselves. From this perspective, the silent era our own cosmos is entering is contrasted by the countless worlds spawning out of the early inflationary era. We are pressed to understand what the correct organization of this late time zoo of metaobservers is, in such an eternally inflating scenario.

As we have already alluded, the second instance of an exponentially expanding universe is where our current expansion is heading. Experimental observations strongly indicate that we are entering a phase dominated by a remarkably small, yet crucially non-vanishing, positive cosmological constant (for a discussion see \cite{Bousso:2007gp}). Its energy density is about $10^{-120}$ times the fourth power of the Planck mass. A universe whose energy density is eventually dominated by a cosmological constant classically asymptotes to a de Sitter space in the far future. Though far from clear, if the current cosmological constant indeed persists for all times this will be the situation future observers will find themselves in. As we already mentioned, observers in such a universe live in a cavity bathed by the Hawking radiation emanating from the cosmic horizon, and are constrained to the observations they make on their finite size lab wall. Virtually all the structure which grew out of the inflationary de Sitter era will be diluted away by the one that awaits. Thus, the question of how to make sense of the cosmic thermal cavity, known as the static patch, becomes vexingly relevant. Equally relevant becomes the question of where such a small cosmological constant could originate from and how to account for the unimaginably vast number of microstates, $\sim 10^{10^{120}}$, coming from the Gibbons-Hawking entropy \cite{Gibbons:1977mu} of the cosmic horizon. From the perspective of an observer in the cosmic cavity, the basic theoretical problem that arises is the lack of a set of sharp observables, due to the lack of any asymptotic accessible boundary.

It is our aim in this article to give an overview of some of the physics of asymptotically de Sitter spacetimes, and some of the attempts to address such questions. Though in no way comprehensive, we try to give an extensive list of references to allow the reader to further explore these matters more readily. We begin by discussing some basic aspects of the geometry of pure de Sitter space and the problem of observables. We continue by discussing the initial value problem for classical four-dimensional general relativity in the presence of a positive cosmological constant. After that, we discuss some notions of perturbative quantum states in a fixed de Sitter background, in both the language of a Fock space and wavefunctionals of field configurations. Then we discuss some known semiclassical results, such as decay rates for nucleation of giant Nariai black holes and the Hartle-Hawking proposal for a wavefunctional. The next section discusses some proposals and speculations for a non-perturbative definition of quantum gravity in an asymptotically de Sitter universe. Finally, we conclude by giving a list of open and potentially tractable questions related to these matters. We should emphasize, this is not a review on inflationary cosmology for which there exist many excellent sources (for example {\cite{Baumann:2009ds}} and references therein). Also, there are many topics we have left untouched even within the context of de Sitter physics, such as the question of infrared divergences which has become an active topic of discussion in recently. This is due to my own lack of expertise on such matters.

\subsection{Geometry of pure de Sitter space}

The pure four-dimensional de Sitter geometry is a maximally symmetric solution to Einstein's equations:
\begin{equation}
\mathcal{G}_{\mu\nu} \equiv R_{\mu\nu} - \frac{1}{2} g_{\mu\nu} R + \Lambda g_{\mu\nu} = 0~,
\end{equation}
with $\Lambda \equiv + 3 / \ell^2$ a positive cosmological constant. The cosmological constant measured in our own universe is given by $\Lambda \sim 10^{-52} \; \text{m}^{-2}$. The de Sitter geometry can be viewed as the induced metric on the four-dimensional hyperboloid:
\begin{equation}
-X_0^2 + \sum_{i=1}^4 X_i^2 = \ell^2~,
\end{equation}
embedded in five-dimensional Minkowski space: $ds^2 = - dX_0^2 + dX_i^2$. The various coordinate patches of de Sitter space cover some or even all of the hyperboloid. In what follows, we will restrict our discussion to four dimensions unless otherwise specified. 

\begin{figure}
\begin{center}
{\includegraphics[height=70mm]{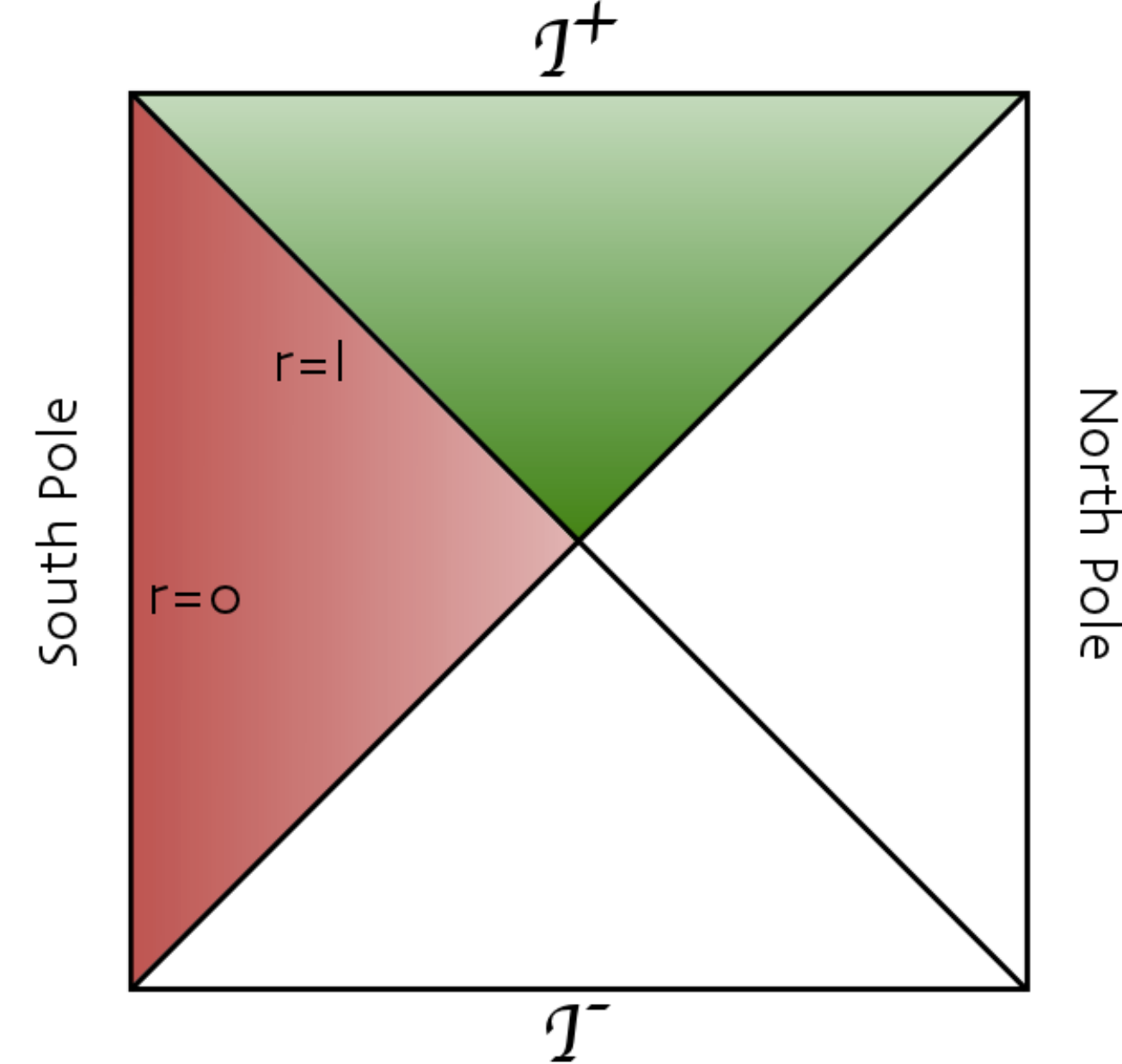} }
   \end{center}\label{penrose} 
   \caption{Penrose diagram of de Sitter space. The full square is given by the global patch (\ref{global}), with each interior point being a two-sphere. Also indicated are the future triangles (green) and the Southern static patches (red), each covering a quarter of the global space.} 
\end{figure}

\begin{figure}
\begin{center}
{\includegraphics[height=63mm]{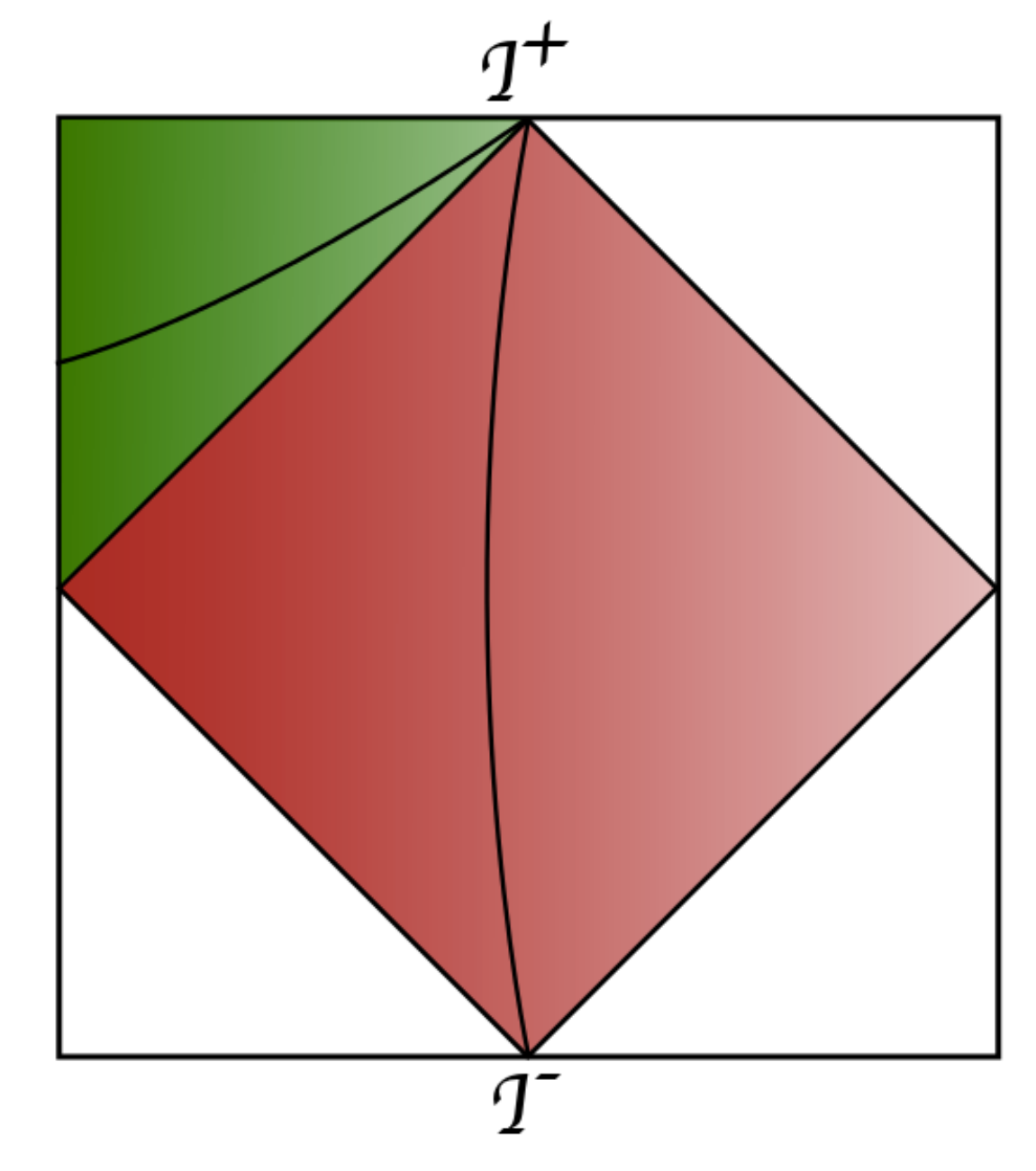} \quad \quad \quad \includegraphics[height=60mm]{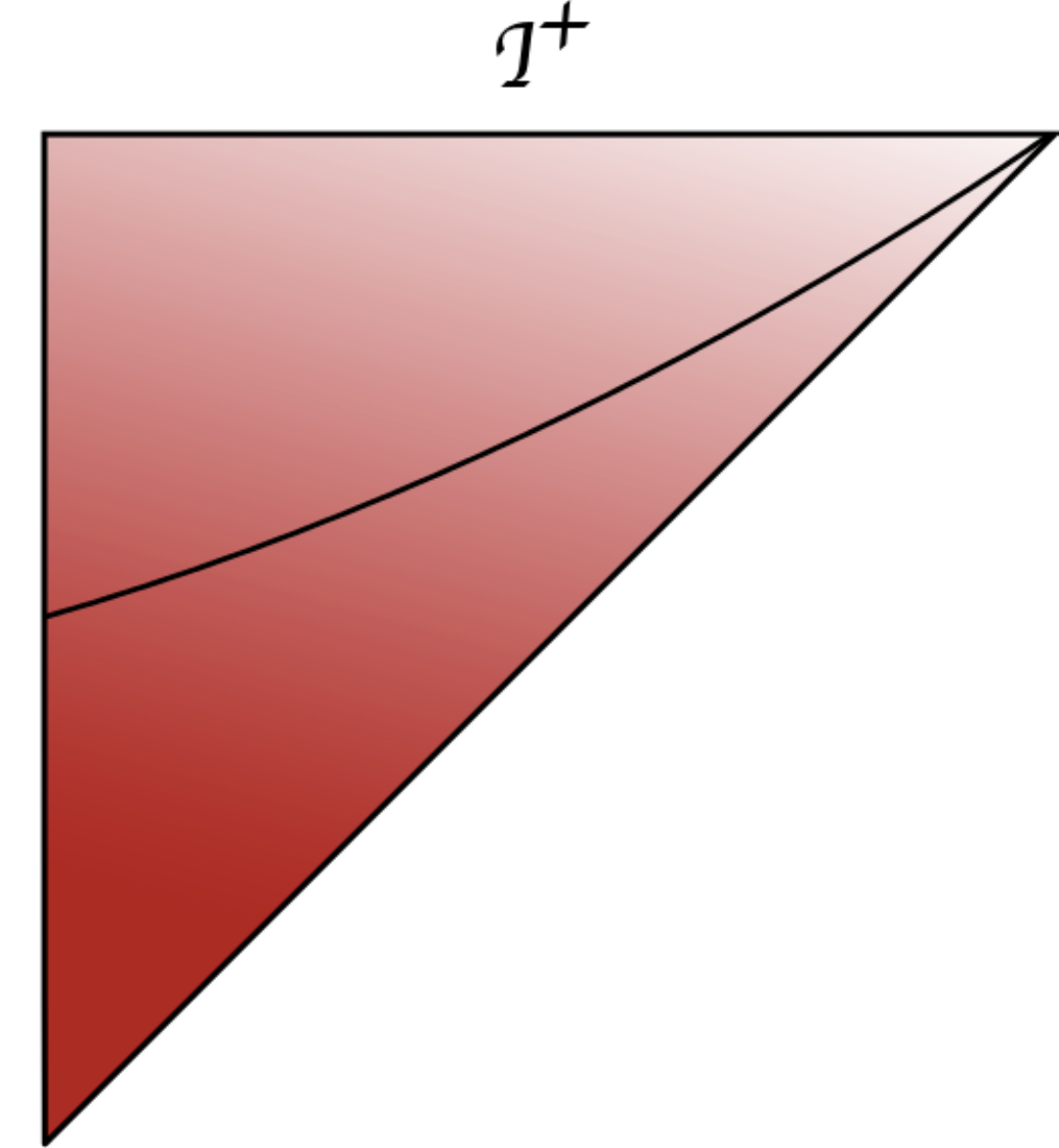} }
   \end{center}
   \caption{Left: Penrose diagram of de Sitter space depicting the de Sitter/de Sitter patch (\ref{dsds}) with a constant spacial dS$_3$ slice (central red diamond) and the hyperbolic patch (\ref{h3ds}) with a constant time $\mathcal{H}_3$ slice (top green corner). Right: Penrose diagram of future directed planar patch (\ref{planar}) containing $\mathcal{I}^+$ and a constant time $\mathbb{R}^3$ slice. }
   \label{fig:planarpenrose}
\end{figure}

The global geometry is described by the metric:
\begin{equation}\label{global}
ds^2 = -d\tau^2 + \ell^2 \cosh^2 \frac{\tau}{\ell} \left( d\psi^2 + \sin^2\psi d\Omega^2_2 \right)~,
\end{equation}
where $\tau \in \mathbb{R}$ and $d\Omega^2_2 \equiv d\theta^2 + \sin^2\theta d\phi^2$ being the round metric on the unit two-sphere. Constant $\tau$ surfaces are three-spheres which shrink from $\mathcal{I}^-$ at $\tau = -\infty$ to $\tau = 0$ and grow from $\tau = 0$ to $\mathcal{I}^+$ at $\tau = +\infty$. The Penrose diagram of global de Sitter is depicted by a square as seen in figure \ref{penrose}. Sometimes, it is more convenient to describe one half of the global geometry using planar spacelike slices:
\begin{equation}\label{planar}
ds^2 = - d T^2 +  \ell^2 e^{2 T / \ell} d\vec{x}^2 = \frac{ \ell^2}{\eta^2}\left( - d\eta^2 + d\vec{x}^2 \right)~,
\end{equation}
with $\vec{x} \in \mathbb{R}^3$ and $T \in \mathbb{R}$. This coordinate patch is depicted in figure \ref{fig:planarpenrose}. The coordinate $\eta = - e^{-T/\ell}$ is known as the conformal time coordinate. The planar metric covers the region of space inside and including the lightcone emanating from a single point at $\mathcal{I}^-$ (which lives at $T \to -\infty$). One can similarly describe the region of space inside and including the past lightcone emanating from a point at $\mathcal{I}^+$ with the same coordinate system. In this case it is convenient to take $T \to -T$ and $\eta \to -\eta$ to keep the time coordinates increasing in the forward time direction.

Neither the global nor the planar geometries can be completely accessed by a single observer. Instead, the region of space accessible to a single observer is known as the the static patch and described by the metric:
\begin{equation}\label{staticpatch}
ds^2 = - \left( 1 - \frac{r^2}{\ell^2} \right) dt^2 + \left( 1 - \frac{r^2}{\ell^2} \right)^{-1} {dr^2} + r^2 d\Omega^2_2~.
\end{equation}
The coordinate ranges are $r \in [0,\ell]$ and $t \in \mathbb{R}$. One may notice that the norm of the $\partial_t$ Killing vector vanishes at $r = \ell$. Indeed, $r = \ell$ is a null surface that surrounds the observer at all times, known as the cosmological horizon. One can also describe the region to the future the cosmological horizon $r \in (\ell,\infty)$ with the above coordinate system, and it describes what is known as the future triangle. A similar statement is true for the region to the past of the cosmological horizon. We can also consider an extension of the static patch that includes the region in a single static patch as well as the future triangle which is smooth across the horizon:
\begin{equation}\label{nullds}
\frac{ds^2}{\ell^2} = \left( - \frac{\rho}{\alpha} + \rho^2 \right)d\tau^2 + d\tau d\rho + \left( 1 - 2 {\alpha}{\rho} \right)^2 d \Omega_2^2~,
\end{equation}
with $\tau \in \mathbb{R}$ and $\rho \in (1/2\alpha,-\infty)$ with $\mathcal{I}^+$ at $\rho \to - \infty$. Constant $\rho$ slices are null rays emanating from the worldline at $\rho = 1/2\alpha$ reaching all the way out to $\mathcal{I}^+$. The horizon is at $\rho = 0$. The patches (\ref{staticpatch}) and (\ref{nullds}) are depicted in figure \ref{penrose}.
We can also describe the hyperbolic patch of de Sitter space described by the metric:
\begin{equation}\label{h3ds}
ds^2 = - d \tilde{t}^2 + \ell^2 \sinh^2 \frac{\tilde{t}}{\ell} \; d\mathcal{H}^2_3~,
\end{equation}
where $\tilde{t} \in \mathbb{R}$ and $d\mathcal{H}_3^2 = d R^2 + \sinh^2 R d\Omega_2^2$ is the standard metric of hyperbolic three-space.\footnote{It is worth mentioning that this geometry (with negative curvature constant $\tilde{t}$ spacelike slices) is the geometry that arises inside a positive $\Lambda$ Coleman-De Luccia bubble nucleated in some false vacuum.} Another patch of de Sitter space is radially foliated by three-dimensional de Sitter space and we call it the de Sitter/de Sitter patch:
\begin{equation}\label{dsds}
ds^2 = d w^2+ \sin^2 \frac{w}{\ell} \left( - d\tilde{\tau}^2 + \ell^2 \cosh^2\frac{\tilde{\tau}}{\ell} d\Omega^2_2 \right)~.
\end{equation} 
The coordinate regions are now $w \in [0,\pi]$ and $\tilde{\tau} \in \mathbb{R}$. Notice that there is a horizon at $w = 0$ and $w = \pi$. The de Sitter/de Sitter and hyperbolic patches are depicted in figure \ref{fig:planarpenrose}. Finally, we can have a patch foliated by $\mathcal{H}_2 \times \mathbb{R}$ slices:
\begin{equation}
ds^2 = - \frac{d\tau^2}{\left(1+ (\tau / \ell)^2 \right)} +\left( 1 + (\tau / \ell)^2 \right) dx^2 + \tau^2 d \mathcal{H}_2^2~,
\end{equation}
with $\tau \in \mathbb{R}$, $x \sim x + \ell$ and $d\mathcal{H}_2^2 = d R^2 + \sinh^2 R d\phi^2$ is the standard metric on hyperbolic two-space. 

The isometries of four-dimensional de Sitter space are given by $SO(4,1)$. This isometry group manifests itself as the conformal group of the three-metric on $\mathcal{I}^\pm$ which may be $\mathbb{R}^3$, $S^2 \times \mathbb{R}$, $\mathcal{H}_3$, $\mathcal{H}_2 \times {S^1}$ or $S^3$ in the case of pure de Sitter. There is no global timelike Killing vector and hence no positive energy theorem, at least in its traditional form. In fact, global slices are compact and consequently conserved (gauge) charges vanish on global slices. One may also consider solutions given by smoothly quotienting the metric on $\mathcal{I}^+$. For example we can quotient $\mathbb{R}^3$ to a three-torus and $\mathcal{H}_3$ to a compact hyperbolic three-manifold. In doing so, one may develop singularities in the bulk where cycles shrink to zero size.

As a final note on the geometry of pure de Sitter space, we mention that its Euclideanization is given by the round metric on the four-sphere. In global coordinates this can be seen by taking $\tau \to i \tau$, in static patch coordinates by taking $t \to i t$, in de Sitter/de Sitter coordinates by taking $\tilde{\tau} \to i \tilde{\tau}$ and in hyperbolic coordinates by taking $d\mathcal{H}_3^2 \to - d\Omega_3^2$ and $\tilde{t} \to i \tilde{t}$. 

\subsection{Observables?}\label{observables}

Perhaps the basic unresolved question for asymptotically de Sitter universes regards the (non)-existence of precise observables (see for example \cite{Banks:2002wr,Bousso:2000md}). In a theory of gravity we must let go of our notions of local observables, as attempting to collect too much information in a confined region of space will eventually cause a gravitational backreaction, and in the most dramatic of cases lead to the formation of a black hole. To avoid this, we typically consider data living at some asymptotic region of space. For example, we could consider the the S-matrix at null infinity of asymptotically flat space or the boundary correlators at the timelike boundary of asymptotically anti-de Sitter spacetimes. De Sitter space has no such asymptotic region given that observers are surrounded by a finite size cosmological horizon. Instead it has asymptotic regions in the infinite future and past. These are infinite spacelike slices which are causally inaccessible to a single observer, save a horizon sized region. Thus, there is a {\it{classical}} uncertainty associated with the measurements of asymptotically de Sitter observers. On the other hand, if the de Sitter length is parametrically large compared to the Planck scale, we should be able to make increasingly precise observations. These observations are naturally made as the data reaches the worldtube of an observatory inside the cosmological horizon.

In this spirit, we begin our discussion with a classical consideration of asymptotically de Sitter universes and the different notions of data used to specify the initial value problem.


\section{de Sitter Space Classical}\label{classics}

The beauty and confusion of de Sitter space already manifests itself at the level of classical gravity. In this section we discuss some considerations of a purely classical four-dimensional de Sitter universe governed by Einstein's equations with a positive cosmological constant. If there is also matter present in our considerations it will always satisfy the null energy condition.

\subsection{Cauchy Problem}

\begin{figure}
\begin{center}
   {\includegraphics[height=32mm]{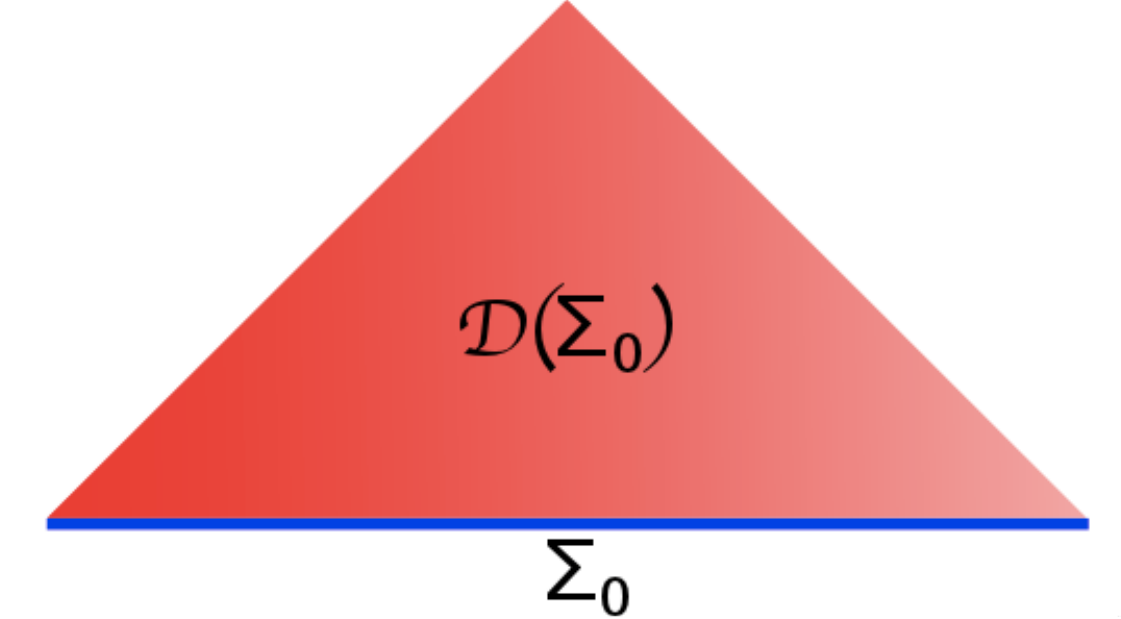} \quad \quad \quad \quad \quad \quad \includegraphics[height=40mm]{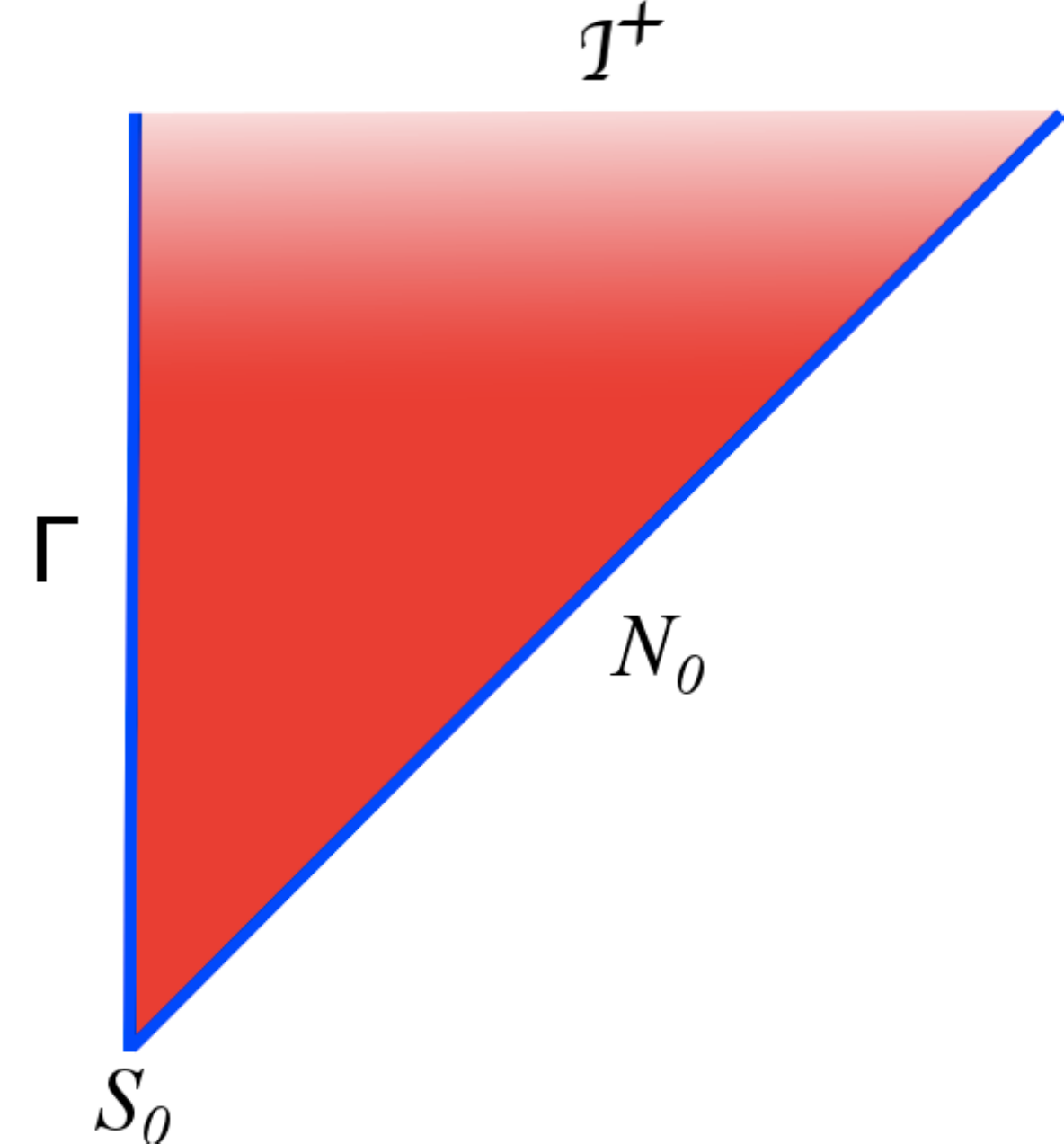}  }
   \end{center}
   \caption{The Cauchy (left) and Tamburino-Winicour (right) boundary vale problems. The data lives on the blue lines and specifies the solution in the shaded region which is either a Cauchy spacelike slice (left) or a worldline with a null line (right).}
   \label{fig:cauchyfig}
\end{figure}

Ordinarily, the way we think about classical gravity is to impose some data, subject to constraints, on a spacelike slice $\Sigma_0$ and evolve it to the future using Einstein's equations $\mathcal{G}_{\mu\nu}$. This is known as the Cauchy problem in general relativity \cite{bruhat,Arnowitt:1962hi,stachel}. The data is given by the induced metric on $\Sigma_0$, denoted by $h_{\mu\nu}$, and the extrinsic curvature of the Cauchy foliation $k_{\mu\nu}$. Let the timelike unit normal to $\Sigma_0$ be given by $n^\mu$, such that $h_{\mu\nu} = g_{\mu\nu}+n_\mu n_\nu$ and $k_{\mu\nu} = \mathcal{L}_n h_{\mu\nu}/2$. One finds that the equations $n_\mu n_\nu \mathcal{G}^{\mu \nu} = n_\mu \mathcal{G}^{\mu i} = 0$ (the index $i$ is tangent to $\Sigma_0$) contain only first derivatives in time and thus act as constraints on $h_{\mu\nu}$ and $k_{\mu\nu}$. These are the momentum and Hamiltonian constraints related to coordinate reparameterizations on $\Sigma_0$ and the absence of a local notion of energy in general relativity due time reparameterization invariance. Evolving the constrained data from $\Sigma_0$ determines the solution in the domain of dependence $\mathcal{D}(\Sigma_0)$ of $\Sigma_0$, given by all points whose causal past lies on $\Sigma_0$ (see figure \ref{fig:cauchyfig}). We can apply the Cauchy problem to Einstein gravity in the presence of a cosmological constant $\Lambda \equiv + 3/\ell^2$. Given that the global spacelike slices are compact, data on an initial spacelike slice $\Sigma_0$ determines the geometry all the way into the future. This may evolve toward a singular solution or to a smooth $\mathcal{I}^+$ depending on the initial data. 

A large class of late time nonlinear solutions\footnote{Though we do not discuss linearized gravity in a fixed de Sitter background in this article, a very complete discussion can be found in {\cite{Kodama:2003kk,Kodama:1985bj}}.} can be expressed as a Fefferman-Graham expansion \cite{Starobinsky:1982mr,fg} in a small parameter $\eta$:
\begin{equation}
\frac{ds^2}{\ell^2} =  -\frac{d\eta^2}{\eta^2} + \frac{1}{\eta^2} \left(  g^{(0)}_{ij} + \eta^2 g^{(2)}_{ij} + \eta^3 g^{(3)}_{ij} + \ldots \right) dx^i dx^j ~.
\end{equation}
The coordinate $\eta$ is the conformal time and the limit $\eta \to 0$ is a late time spacelike slice that tends to $\mathcal{I}^+$. We have fixed the synchronous gauge $g_{\eta\eta} = -\eta^{-2}$ and $g_{\eta i} = 0$. The boundary data is given by the three-metric $g^{(0)}_{ij}$ on $\mathcal{I}^+$ and a transverse-traceless tensor $g^{(3)}_{ij}$, $\text{tr}_{g^{(0)}} g^{(3)}_{ij} = \nabla^i_{g^{(0)}} g^{(3)}_{ij} = 0$. The coefficients of all other powers of $\eta$ are completely determined by $g^{(0)}$ and $g^{(3)}$.\footnote{For example $g^{(2)}_{ij} = R_{ij}[g^{(0)}] - R[g^{(0)}] g^{(0)}_{ij}/4$, which is the Schouten tensor of $g^{(0)}_{ij}$ in three-dimensions.} One can use diffeomorphisms tangent to $\mathcal{I}^+$ to kill three components of $g^{(0)}_{ij}$ and one of the four-dimensional ones to fix its determinant, leaving two degrees of freedom in $g^{(0)}_{ij}$. The transverse-traceless property of $g^{(3)}_{ij}$ leaves two degrees of freedom in $g^{(3)}_{ij}$ as well. Together, $\left(g^{(0)}_{ij},g^{(3)}_{ij}\right)$ form the boundary data of an asymptotically de Sitter universe. More precisely, it is the conformal class $\left(g^{(0)}_{ij}, g^{(3)}_{ij} \right) \sim \left(\Omega^2 g^{(0)}_{ij}, \Omega^{-1} g^{(3)}_{ij}\right)$, for some smooth non-zero function $\Omega(x^i)$, that constitutes the boundary data. 

An important mathematical theorem due to Friedrich \cite{friedrich}, is that the above Cauchy problem is well posed. Given the data $\left(g^{(0)}_{ij}, g^{(3)}_{ij} \right)$ at $\mathcal{I}^+$ and assuming the Cauchy slices are compact (though not necessarily topologically three-spheres) there is a unique extension of the solution into the bulk realizing the data at $\mathcal{I}^+$. Furthermore, small deviations from this data result in small variations of the extended bulk solution. The result was extended to all even dimensions by Anderson \cite{Anderson:2004ir}. From the other end, one can ask what class of initial conditions leads to an asymptotically (locally) de Sitter universe in the future. It has been proposed in what has come to  be known as the {\it{cosmic no hair theorem}} \cite{Gibbons:1977mu,Hawking:1981fz} that (almost) all expanding solutions of the Einstein equations in the presence of a positive $\Lambda$ classically evolve to a locally de Sitter universe in the far future. This proposal has been verified for several classes of geometries \cite{Wald:1983ky,Jensen:1986nf,Barrow:1987ia,VargasMoniz:1992za},  though a small class of counterexamples is also known.

A large class of $\left(g^{(0)}_{ij}, g^{(3)}_{ij} \right)$ at $\mathcal{I}^+$ will contain a big bang type singularity in the past. Similarly, given that the global geometry is contracting if we begin at $\mathcal{I}^-$, a large class of data at $\mathcal{I}^-$ will never evolve into asymptotically de Sitter space due to the formation of a big crunch. There exist results of interest relating the topology and curvature of the geometry at $\mathcal{I}^{\pm}$ with the existence of bulk singularities \cite{Andersson:2002nr}. For instance, in four dimensions in order for there to be a geometry smoothly connecting $\mathcal{I}^-$ to $\mathcal{I}^+$ the (compact) Cauchy surfaces must have a finite fundamental group, like $S^3$ or quotients thereof. Also, if the conformal class of metrics on $\mathcal{I}^-$ contains a metric with constant negative scalar curvature then all timelike geodesics are future incomplete \cite{Andersson:2002nr}.

\subsection{Null-Timelike Initial Value Problem}

Recall that the global geometry of asymptotically de Sitter spacetimes cannot be completely accessed by a local observer. From a local perspective, such as an observer restricted to a lab confined to live inside some finite size worldtube, a large part of the initial data on the $\Sigma_0$ will remain out of causal contact and never be observed.\footnote{We are assuming here that there is no exit from inflation, i.e. our space is asymptotically de Sitter everywhere. Observers that do exit from inflation may be able to access a larger part of $\Sigma_0$.} We can picture this by drawing the past null cone of the lab wall which has reached $\mathcal{I}^+$. In particular, we can partition $\Sigma_0$ into its observable part $\Sigma_{obs}$ and its unobservable part $\Sigma_u$. The physical relevance of data that is forever unobservable is questionable. In any event, we are prompted to consider a more `observer friendly' boundary value problem. This was addressed for vanishing $\Lambda$ by Tamburino and Winicour \cite{tw}. These authors considered the problem of data on a worldtube $\Gamma$ of some finite thickness and a null surface $N_0$ emanating toward the future from some initial spacelike slice of $\Gamma$ which we may denote by $S_0$ (see figure \ref{fig:cauchyfig}). A natural coordinate system for this problem, first considered in \cite{bondi,sachs}, is:
\begin{equation}
ds^2  = - g_{uu} du^2 - 2 \left( g_{u r} dr + g_{u a} dx^a \right) du + r^2 \bar{g}_{ab} dx^a dx^b~, \quad a,b = 1,2~.
\end{equation}
The determinant of $\bar{g}_{ab}$ is fixed to be: $\det \bar{g}_{ab} = f(x^a)^2$ where $f(x^a)$ is a function of the $x^a$ only. The $u$-coordinate is a timelike coordinate on the worldline and the $r$-coordinate describes the position on the null surface emanating from $\Gamma$ at some given $u$. The $x^a$-coordinates describe the position on the two-sphere at constant $r$ and $u$. 

It was shown \cite{tw} (see also \cite{dinverno}) that knowing $\bar{g}_{ab}$ on $\Gamma$ as a function of $u$ and $x^a$, $\bar{g}_{ab}$ on $N_0$ as a function of $r$ and $x^a$ and $g_{uu}$, $g_{ua,r}$ on $S_0$ (as functions of $x^a$ only) allows us to determine the solution to Einstein's equation in the region between $\Gamma$ and $N_0$. Though this problem has not been addressed in the context of Einstein's equations with positive $\Lambda$ and it would be very interesting to do so. In particular, one might want to understand the relation between the data on $\left(\Gamma, N_0, S_0 \right)$ and the Fefferman-Graham data $\left(g^{(0)}_{ij}, g^{(3)}_{ij}\right)$ at $\mathcal{I^+}$. It is also worth mentioning that the consideration of data on a timelike surface has played a role in asymptotically anti-de Sitter spacetimes. In this case, one specifies sources on the timelike boundary and studies their response by solving for the bulk.


\subsection{Bondi Problem and Sach's Double Null Problem}\label{sachsdoublenull}

\begin{figure}
\begin{center}
 {\includegraphics[height=35mm]{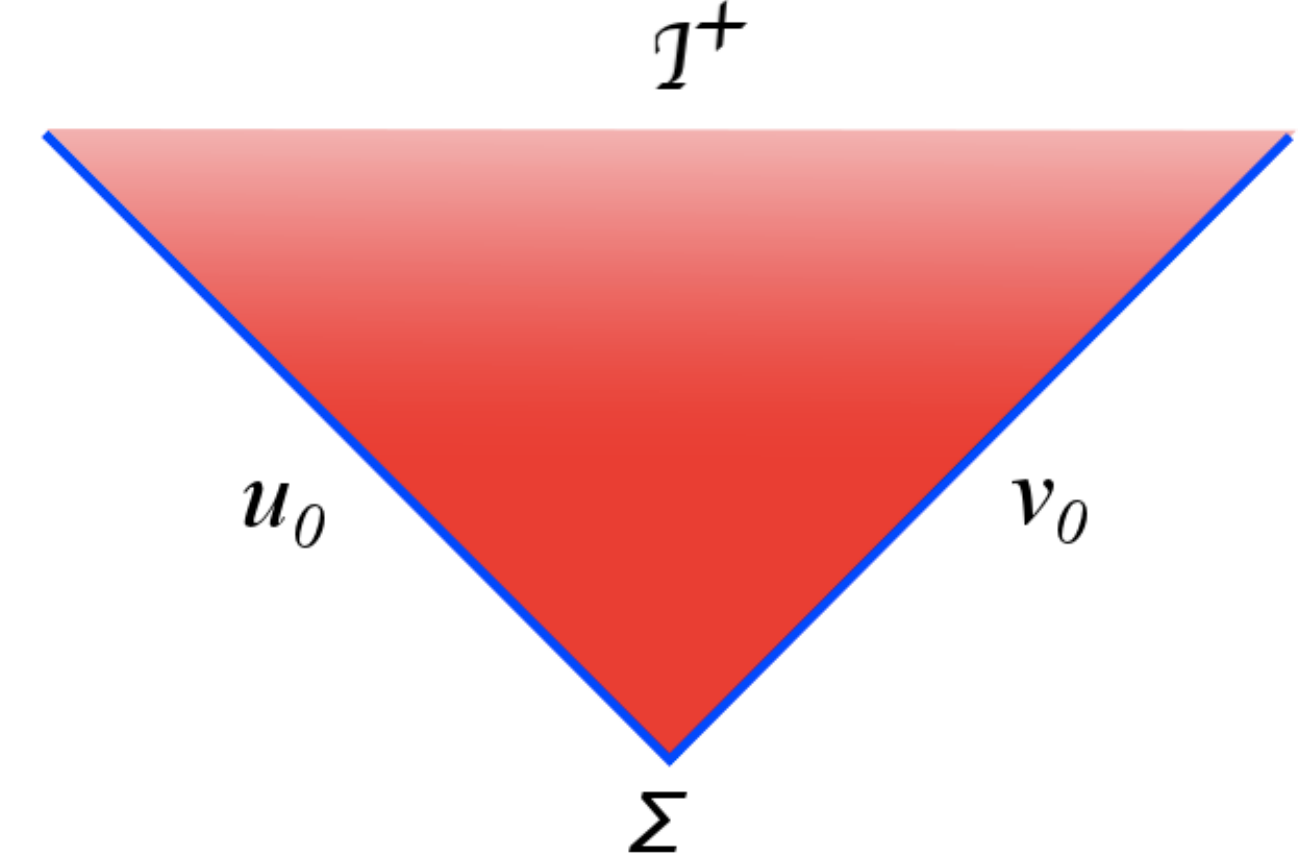}  \quad \quad \quad \quad \quad \quad \includegraphics[height=40mm]{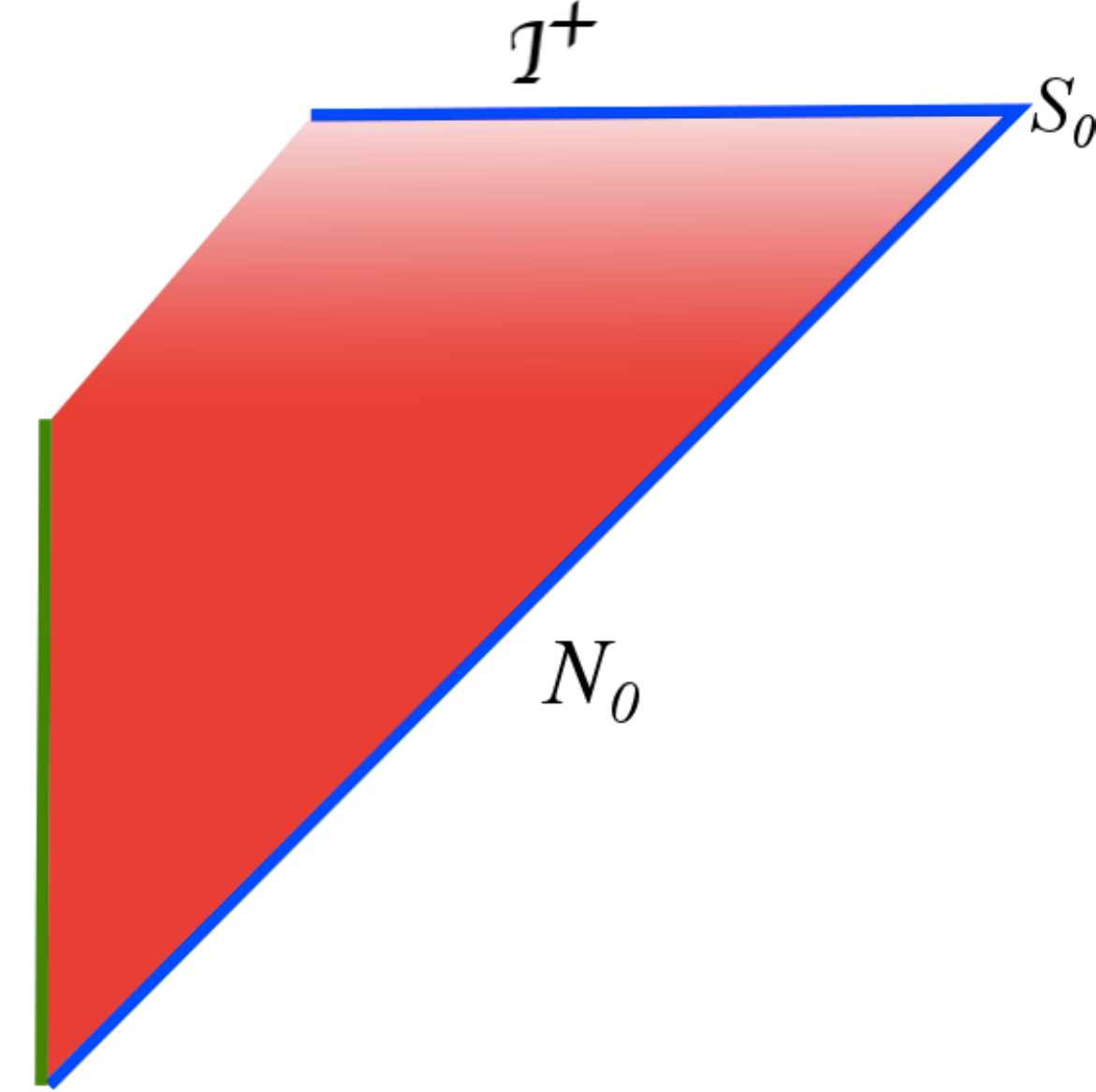} }
 \end{center}
   \caption{The Sachs double null (left) and Bondi (right) boundary vale problems. The data lives on the blue lines and specifies the solution in the shaded (red) region. The green line indicates some localized radiating source in the Bondi problem.}
   \label{fig:bondifig}
\end{figure}

For the sake of completeness we mention two other boundary value problems that have been considered when studying Einstein's equations with vanishing cosmological constant. These are depicted in figure \ref{fig:bondifig}.

The first is the Bondi problem \cite{bondi}, which was considered in \cite{chrusciel} for four dimensional de Sitter space. The worldline data on $\Gamma$ we previously considered is pushed all the way to $\mathcal{I}^+$.
This data, which lives on an interval of $\mathcal{I}^+$, in addition to the data on a null slice $N_0$ emanating from a localized radiating source to $\mathcal{I}^+$ as well as data on a two-sphere $S^0$ at $\mathcal{I}^+$ specifies a solution.

The other setup, considered by Sachs \cite{sachsdn}, specifies data on two null slices intersecting at a common two-surface $\Sigma$. For the double null problem (with $\Lambda=0$), one builds a line element of the form:
\begin{equation}
ds^2 = -e^{2q} du dv + e^{2h}\bar{g}_{ab} \left( d x^a + C^a du \right)(  d x^b + C^b du )~,
\end{equation}
with $\det{\bar{g}_{ab}} = 1$, $C^a = 0$ on the initial null slice $v = v_0$ and $q = 0$ on the initial null slices $v = v_0$, $u = u_0$. The data $\bar{g}_{ab}(x^a, u)$ on $v=v_0$ and $\bar{g}_{ab}(x^a, v)$ on $u=u_0$ in addition to $h$, $C_{a,v}$, $h_{,u}$ and $h_{,v}$ on $\Sigma$ determine a solution to the future of the null slices. It can be shown that an analogous setup exists in the presence of a non-vanishing, positive $\Lambda$ \cite{diogim}. We should emphasize that constructing the geometry in the future triangle (outside the observer's horizon) from data on two null slices allows us to keep track of how this geometry is related to the data of the Northern and Southern static patches in figure \ref{penrose}. For instance, it allows is to consider variations of the data from a single static patch on the $v=v_0$ null slice while maintaining the data on the $u = u_0$ null slice fixed.

\subsection{Asymptotic Symmetries}

The notion of symmetry in theories of gravity is subtle. Though some special solutions have an isometry group associated to them, generically this is not the case since the metric is sensitive to all surrounding matter and the symmetries are destroyed. A useful notion that appears in the context of classical general relativity is that of asymptotic symmetries. These are the set of diffeomorphisms $\xi^\mu$, subject to some (physically motivated) boundary condition, that alter the physical boundary data. In certain circumstances, one can associate to each of the asymptotic symmetries a charge $Q_{\xi}$ generating the symmetry and study the algebra of the charges. Choosing the appropriate boundary condition is somewhat context dependent. A well known example is given by the symmetry group of asymptotically flat spacetimes. This was defined in the seminal works of Bondi, Metzner \cite{bondi} and Sachs \cite{Sachs:1962zza}. They found that the asymptotic symmetries of asymptotically flat spaces comprise an \emph{infinite} dimensional group, the BMS group, containing the (finite dimensional) Poincare symmetry as a subgroup. 

The asymptotic symmetry group of four-dimensional de Sitter space has been analogously studied \cite{chrusciel,Anninos:2010zf,Anninos:2011jp,Kelly:2012zc}. In the global case it was found that for boundary conditions allowing variations of the conformal metric $g^{(0)}_{ij}$ on $\mathcal{I}^+$ the asymptotic symmetries comprise the full set of diffeomorphisms tangent to $\mathcal{I}^+$. Given a bulk spacelike three-slice $\Sigma$ intersecting $\mathcal{I}^+$ at some two-slice $\partial \Sigma$, the charges associated to the $\xi$ are found to be\footnote{Notice that the charges are given solely by a boundary term which would not be present if the slice did not cut $\mathcal{I}^+$ at some $\partial \Sigma$. Indeed on a compact global three-slice $\Sigma$ the charges above would vanish.} the Brown-York charges \cite{Brown:1992br,Abbott:1981ff,Balasubramanian:2001nb,Compere:2008us,Wald:1999wa,de Haro:2000xn,Henningson:1998gx}
\begin{equation}
Q_\xi [\partial\Sigma] = -\frac{3\ell}{16\pi G} \int_{\partial \Sigma} d^2 x \sqrt{\sigma} n^i \xi^j g^{(3)}_{ij}~,
\end{equation}
where $\sigma$ is the determinant of the induced metric on $\partial \Sigma$ and $n^i$ is normal to $\partial \Sigma$. Furthermore, the flux between bulk two slices intersecting $\mathcal{I}^+$ at $\partial\Sigma_1$ and $\partial\Sigma_2$ is given by \cite{Anninos:2010zf}:
\begin{equation}
Q_\xi [\partial\Sigma_2]  - Q_\xi [\partial\Sigma_1] = -\frac{3\ell}{32\pi G} \int_{\mathcal{B}_{12}} g^{(3)ij} \mathcal{L}_\xi {g^{(0)}_{ij}}\sqrt{g^{(0)}} d^3x~.
\end{equation}
where $\mathcal{B}_{12}$ is the region on $\mathcal{I}^+$ between $\partial\Sigma_1$ and $\partial\Sigma_2$.

As we mentioned before, for a local (eternally de Sitter) observer who cannot access the full data on a spacelike slice $\Sigma_0$, it is unclear what the physical meaning of the inaccessible data is. On the other hand, along with the wordline data of the observer herself, this inaccessible data determines the geometry outside the cosmological horizon, i.e. in the future triangle. Thus, one may ponder on other interesting ways to fix the data on $\Sigma_u$. One motivation may come from requiring that the asymptotic symmetries comprise the isometry group of pure de Sitter space, $SO(4,1)$, in analogy to the case of anti-de Sitter space. In order to do so, one could tune the data on $\Sigma_u$ such that for any variation of data on $\Gamma$ and the observable part of $N_0$, the variations of $g^{(0)}_{ij}$ vanish. Though such an initial value problem severely violates causality, this violation is classically inaccessible to the local observer. Evidence that this is possible was given at the linearized level in \cite{Anninos:2011jp}. Another possibility is to consider freezing the unobservable data, such that the variations of data reaching $\mathcal{I}^+$ are encoded fully by variations of data on $\Gamma$ and the observable part of $N_0$ \cite{Anninos:2011zn}.

\subsection{Cosmic Fluids}

Let us now consider again the local observer and her worldtube. If we push the size of our worldtube to approach the de Sitter radius $\ell$, we are forced to study the dynamics of metric deformations very near the cosmological horizon. The problem mimics similar analyses in the context of the near horizon regions of black hole geometries \cite{Price:1986yy,Bhattacharyya:2008kq,Bhattacharyya:2008jc,Bredberg:2011jq,Bredberg:2011xw,Eling:2009pb} beginning with the seminal work of Damour \cite{damour}. The basic idea is that solutions to Einstein's equations in the near horizon limit are organized in the form of solutions to a Navier-Stokes equation. This is related to the fact that near horizon regions encode the deep infrared behavior of fields (due to an increasing redshift factor) and the Navier-Stokes equation encodes the deep infrared physics of a rather universal set of systems. It was shown in \cite{Anninos:2011zn} that the dynamics of non-linear metric deformations very near the cosmological horizon are indeed governed by the incompressible (non-relativistic) Navier-Stokes equation. The radial direction emanating away from the timelike surface is $\rho$ and the time coordinate is $\tau$. Constant $\tau$ and $\rho$ surfaces are two-spheres parametrized by $\Omega = (\theta,\phi)$.
The deformation is expanded in a dimensionless near horizon parameter $\alpha$ which is taken to be small, and parametrized by $v^i$, $P$  and $\phi_i^{\left(\alpha\right)}$ which are functions of $(\tau,\Omega^i)$ only. 
As Dirichlet boundary conditions on the $\rho = 1$ timelike hypersurface (which in the $\alpha \to 0$ limit is parametrically close to the horizon) one requires the perturbations to preserve the induced metric on the hypersurface $\rho=1$:
\begin{equation}
ds_{3d}^2= \left(-\frac{1}{\alpha} + 1\right)d\tau^2 + \left(1-2 \alpha\right)^2 d\Omega^2_2~,
\end{equation}
 up to a conformal factor:
\begin{equation}
1 + 2 \alpha P + \mathcal{O}(\alpha^2)~.
\end{equation}
Then, imposing the Einstein equations with $\Lambda > 0$ and taking the limit $\alpha \to 0$, $v^i$ and $P$ are found to obey:
\begin{equation}\label{navierstokesfluid}
\partial_\tau v^i + \nabla_{S^2}^i P + v_j \nabla^j_{S^2} v^i - \nu \left( \nabla^2_{S^2} v^i + R^{i}_{j} v^j \right) = 0~, \quad \nabla_{S^2}^i v_i = 0~,
\end{equation}
to leading order in $\alpha$. As mentioned, this is nothing more than the incompressible Navier-Stokes equation with $\nu = 1 + \mathcal{O}(\alpha)$. 

\subsection{Quasinormal Modes}\label{qnmsec}

As a final note we discuss the quasinormal mode spectrum of de Sitter space. Quasinormal modes are a particular set of perturbative modes that encode the dissipative nature or `ringing' of a given background \cite{Kokkotas:1999bd}. The recipe to compute them is given by solving the linearized equations of motion of some matter field, such as the graviton, and demanding that the solutions obey specific properties. In the case of an asymptotically flat Schwarzschild solution, it is demanded that the modes are purely outgoing at future null infinity and purely infalling into the black hole's future horizon. In the de Sitter case \cite{LopezOrtega:2006my}, working with the static patch coordinates (\ref{staticpatch}), we impose no incoming flux from the past cosmological horizon and that the modes be regular near the worldline at $r = 0$. The idea is to capture the ringing of the spacetime by sending an isolated pulse from the worldline and study the behavior at late times. 

As an example consider a scalar field $\Phi(x)$ of mass $m$. Expanding the field in a Fourier basis $\Phi(x) = e^{-i \omega t} Y_{l m}(\theta,\phi) R_{l m}^{ \omega} (r)$, we find that these boundary conditions restrict the $\omega$'s to a discrete set:
\begin{equation}\label{quasinormalfreq}
\omega^\pm_n \ell = - i \left( 2 n + l + \frac{3}{2} \pm \sqrt{\frac{9}{4} - m^2 \ell^2}\right)~.
\end{equation}
Similar results hold for more general fields. Notice that for sufficiently light fields, the de Sitter quasinormal mode spectrum is pure imaginary. This is in contrast to the quasinormal mode spectrum of the Schwarszchild black hole \cite{Bachelot:1993dp,Motl:2002hd,Motl:2003cd} or asymptotically AdS black holes \cite{Horowitz:1999jd,Birmingham:2001pj} which typically have a real part which is some function of the momenta. Given that $\omega_n^\pm$ have a negative imaginary part, the de Sitter quasinormal modes decay exponentially at late times.
 
\section{de Sitter dressed in Black}

In this section we briefly discuss a particularly interesting non-linear solution to Einstein's equations with positive $\Lambda$: the Schwarzschild-de Sitter geometry. The metric for the non-rotating case is given by:
\begin{equation}\label{schw}
{ds^2} = -\left( 1 - \frac{r^2}{\ell^2} - \frac{2M}{r} \right)dt^2 + {\left( 1 - \frac{r^2}{\ell^2} - \frac{2M}{r} \right)^{-1}}{dr^2} + r^2 d\Omega^2_2~.
\end{equation}
The $g_{tt}$ component of the metric has two positive real zeros, $r_c(M)$ and $r_{+}(M)$ with $r_c > r_{+}$ which are the cosmological and black hole horizons. It has been shown in \cite{Kodama:2003jz} that the Schwarzschild-de Sitter geometry is perturbatively stable with respect to gravitational perturbations. As the parameter $M > 0$ increases, $r_c$ and $r_{+}$ tend to each other and eventually meet at a critical mass $M_c = \ell/3\sqrt{3}$. This is known as the Nariai limit. For $M > M_c$ one finds a cosmological solution with no horizons reaching all the way to $\mathcal{I}^+$ with a spacelike singularity at $r = 0$, where $r$ is now a time  coordinate. Indeed, for $M>M_c$ one can identify $t \in \mathbb{R}$ without introducing closed timelike curves. For $M < 0$, one finds only one horizon and a naked timelike singularity at $r = 0$. It is important to note that there are {\it no} planar black hole solutions in a theory of gravity with positive $\Lambda$, in stark contrast to the AdS case. 

\begin{figure}
\begin{center}
   \includegraphics[height=70mm]{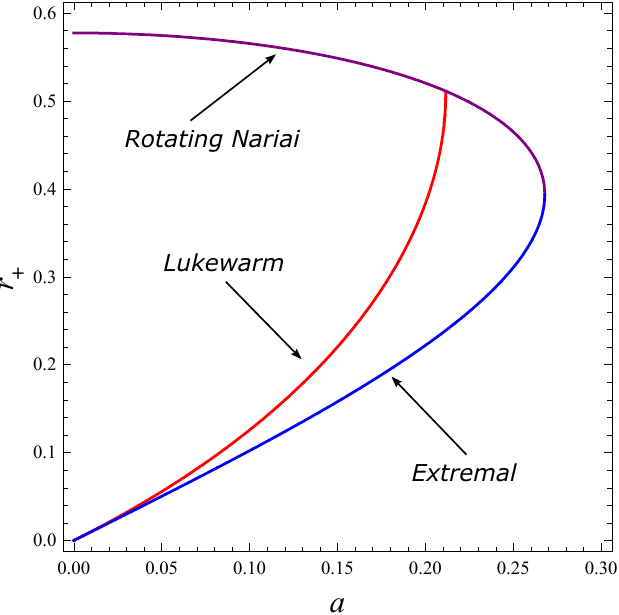}
   \end{center}
   \caption{Region in the $(r_+, a)$-plane allowing for smooth black hole solutions. The static patch geometry lives at the origin.}\label{phasespace}
\end{figure}

One can generalize to the case of rotating black holes as well which are specified by the mass parameter $M$ and the spin parameter $a$. The horizons are now given by the zeroes of:
\begin{equation}
\Delta_r = (r^2 + a^2) \left( 1 - \frac{r^2}{\ell^2} \right) - 2 M r~.
\end{equation}
The geometry of rotating black holes in de Sitter is discussed for example in \cite{Booth:1998gf,Anninos:2009yc}. The function $\Delta_r$ has four real roots when $M > 0$. These correspond to the cosmological horizon $r_c$, the outer and inner black hole horizons $0 < r_\pm < r_c$ and a horizon $r_n < 0$ living behind the ring singularity at $r = 0$. When the outer and inner black hole horizons coincide, we have an extremally rotating black hole. For a special function of $M(a)$ one finds that the surface gravity of the black hole and cosmological horizons coincide and the black hole is known as lukewarm. When the cosmological and outer black hole horizons coincide we have a rotating Nariai geometry \cite{Booth:1998gf}. In figure \ref{phasespace} we show the allowed phase space of non-singular de Sitter black holes as a function of $r_+$ and the spin parameter $a$. Notice that one cannot make arbitrarily large black holes for any $M$ and $a$ and in fact the space with the largest cosmological horizon is pure de Sitter space itself! These are strikingly different features from say anti-de Sitter or flat space black holes.

\subsection{Nariai Geometry}

\begin{figure}
\begin{center}
   \includegraphics[height=55mm]{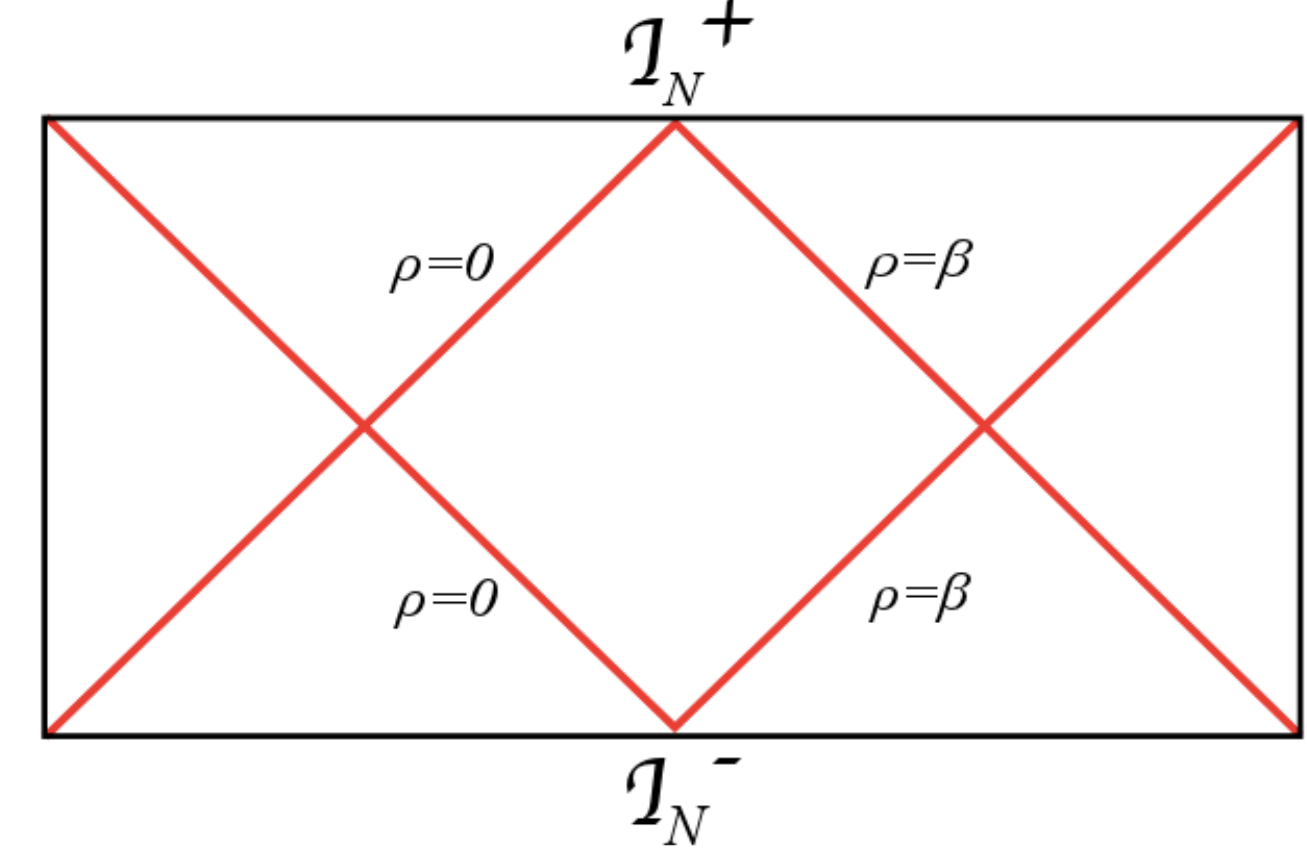}
   \end{center}
   \caption{Penrose diagram of dS$_2$.}
   \label{fig:ds2}
\end{figure}

In the case $r_c \to r_{+}$ we can in fact take a near horizon limit, zooming into the region between the horizons. This is accomplished by introducing:
\begin{equation}
\tau = \lambda t / r_c~, \quad \rho = \frac{r-r_{+}}{r_c \lambda}~, \quad \beta = \frac{r_c - r_{+}}{r_c \lambda}~,
\end{equation}
and taking the limit $\lambda \to 0$ while keeping $\beta$ fixed. This leads to the Nariai metric \cite{nariai}:
\begin{equation}\label{nariaimetric}
{ds^2} = \frac{\ell^2}{3}\left(-d\tau^2 \rho(\rho - \beta) + \frac{d\rho^2}{\rho(\rho - \beta)} + d\Omega^2_2 \right)~,
\end{equation}
which is nothing more than the dS$_2 \times S^2$ geometry. The Penrose diagram of dS$_2$ is depicted in figure \ref{fig:ds2}. The original black hole horizon now lives at $\rho = 0$ and the cosmological horizon lives at $\rho = \beta$. The quasinormal modes of the Nariai black hole in de Sitter space were considered in \cite{Cardoso:2003sw} and are found to be:
\begin{equation}
\omega_n \ell = \frac{\beta}{6} \left( - \left( n + \frac{1}{2} \right) i + \sqrt{(l+2)(l-1)-\frac{1}{4}} \right)~,
\end{equation}
where $l$ is the angular momentum on the $S^2$ and the frequency $\omega_n$ is measured with respect to the $\tau$-coordinate in (\ref{nariaimetric}). When geodesically completing the space to:
\begin{equation}\label{globnariai}
ds^2 = \frac{1}{3} \left( - d\tau^2 + \cosh^2 \frac{\tau}{\ell} d\psi^2 + \ell^2 d\Omega_2^2 \right)~,
\end{equation}
a new future/past boundary $\mathcal{I}^\pm_N$ develops at $\tau \to \pm \infty$. Notice that (\ref{globnariai}) is {\it not} an asymptotically dS$_4$ universe. On the other hand, it is unknown what class of initial Cauchy data preserves the asymptotic structure of the global Nariai universe and it may be extremely limited \cite{Beyer:2009vu,Beyer:2009vv}.

Had we also included angular momentum we would have found a similar near horizon limit whose geometry is given by the one-parameter family of rotating Nariai metrics. In global coordinates:
\begin{equation}\label{rotnariai}
ds^2 = \Gamma(\theta) \left[ -d \tau^2 +\cosh^2 \frac{\tau}{\ell} d \psi^2 + \alpha(\theta) d\theta^2 \right] + \gamma(\theta) \left( d\phi - k \sinh \frac{\tau}{\ell}  d\psi \right)^2~.
\end{equation}
The functions $\Gamma(\theta)$, $\alpha(\theta)$ and $\gamma(\theta)$ can be found in \cite{Booth:1998gf,Anninos:2009yc}. Geometrically, the space is an $S^2$ fibered over a dS$_2$ base space and there exists a future and past boundary $\mathcal{I}^\pm_{RN}$ at $\tau \to \pm \infty$. In this limit, $a$ and $M$ are related to $r_c$ as:
\begin{equation}
a^2 = \frac{r_c^2 \left( 1 - 3 r_c^2/ \ell^2 \right)}{1+ r_c^2/\ell^2}~, \quad M = \frac{r_c \left( 1 - r^2_c / \ell^2 \right)^2}{1 + r_c^2 / \ell^2}~.
\end{equation}
It is worth mentioning that the asymptotic symmetry group at $\mathcal{I}^+_{RN}$ of the rotating Nariai geometry is the (infinite-dimensional) Virasoro algebra with central charge \cite{Anninos:2009yc}:
\begin{equation}
c_L = \frac{12 r_c^2 \sqrt{\left(1 - 3 r_c^2/\ell^2\right)\left(1+r_c^2/\ell^2\right)}}{-1+ 6 r_c^2 / \ell^2 + 3 r_c^4 / \ell^4}~.
\end{equation}
The Virasoro algebra is the symmetry group of two-dimensional conformal field theories and may suggest that the language of 2d CFT's is intricately connected to the rotating Nariai geometry, perhaps holographically. Further evidence for this idea was pursued through an extensive study of quantum fields in the rotating Nariai background \cite{Anninos:2010gh}. We could also consider adding a $U(1)$ gauge symmetry and study charged black holes in de Sitter space. Quite remarkably, in this case a set of time dependent multi-black hole solutions have been discovered by Kastor and Traschen \cite{Kastor:1992nn}.

Finally, we would like to make a brief comment about a set of black hole solutions to Einstein gravity with a {\it negative} cosmological constant, known as topological AdS$_4$ black holes. These black holes have a horizon with a $\mathcal{H}_2$ geometry and can have lower energy than the AdS$_4$ vacuum with $\mathcal{H}_2$ slicing. However, their energy is bounded from below by the critical hyperbolic black hole which has a near horizon geometry given by AdS$_2 \times \mathcal{H}^2$. The set of topological AdS$_4$ black holes with energy lower than the AdS$_4$ vacuum is a direct analytic continuation of the de Sitter black holes and in particular, the AdS$_2 \times \mathcal{H}_2$ solution is an analytic continuation of the dS$_2 \times S^2$ Nariai solution.


\section{de Sitter Space (Slightly) Quantum}

In this section we discuss the quantization of a non-interacting massive scalar field in a fixed de Sitter background. When working in a flat space background with a global timelike Killing vector, one finds it rather straightforward to define what is meant by energy. This comfort quickly dissipates in curved backgrounds and almost completely disappears in a time-dependent background where there is no global timelike Killing vector. Such is the case of de Sitter space. What do we mean by the `vacuum state' of a quantum field?\footnote{A wonderful overview of quantum field theory in curved backgrounds is given by \cite{birrel}.}

\subsection{Bunch-Davies State}

Let us for simplicity consider a free scalar field $\Phi(x) = \eta \varphi(\eta)e^{i \vec{k}\cdot\vec{x}}$ with mass $m$ in a fixed de Sitter background with planar slices (\ref{planar}). The wave equation is given by:
\begin{equation}\label{eom}
\left( \eta^2 \partial_\eta \eta^{-2} \partial_\eta +  k^2 \right)\eta \varphi = - \eta^{-1} \ell^2 m^2 \varphi~.
\end{equation}
There are two-linearly independent solutions to this equation given by:
\begin{equation}
\varphi_1 = \frac{1}{2}(\pi\eta)^{1/2} J_\nu (k\eta)~, \quad \varphi_2 = \frac{1}{2}(\pi\eta)^{1/2} Y_\nu (k\eta)~.
\end{equation}
We have normalized with respect to the Klein-Gordon inner product,
\begin{equation}
(\varphi_n,\varphi_m) = -i \int_\Sigma d^3 x \sqrt{h} n^\mu \left(\eta \varphi_n \overleftrightarrow{\partial_\mu}( \eta \varphi_m^* ) \right) = \delta_{nm}~,
\end{equation}
where $\Sigma$ is a spacelike three-slice with induced metric $h_{ij}$ and future directed norm $n^\mu$. The functions $J_\nu(z)$ and $Y_\nu(z)$ are Bessel functions of the first and second kind and $\nu \equiv \sqrt{9/4-m^2\ell^2}$. One can also solve the equation of motion near $\mathcal{I}^+$, i.e. $\eta \to 0$, to find the following asymptotic form:
\begin{equation}
\varphi_\pm \sim (-\eta)^{1/2 \pm \nu}~.
\end{equation}
Notice that $\varphi_+$ becomes an oscillatory mode with positive frequency $|\nu|$ for $m\ell>3/2$. Finally, we can examine the modes in the region $k |\eta| \to \infty$ residing well within the cosmological horizon. We find that the combination:
\begin{equation}
\lim_{k\eta\to\infty} \left(\varphi_1 - i \varphi_2\right) \to \frac{1}{\sqrt{2k}} e^{-ik\eta}~,
\end{equation}
is equivalent to a positive frequency mode in Minkowski space with respect to the canonical vacuum state.

Positive frequency modes in the Bunch-Davies (or Euclidean) vacuum state $|E\rangle$ are those which become the positive frequency modes in Minkowski space upon taking the limit $k|\eta| \to \infty$ \cite{Bunch:1978yq} (see also \cite{Sasaki:1994yt}). These are given precisely by $\varphi_E \equiv \varphi_1 - i \varphi_2$. We can thus expand our quantum field in terms of the creation and annihilation operators of $|E\rangle$:
\begin{equation}\label{Emodes}
\hat{\varphi}_E (\eta,\vec{x}) = \sum_{k>0} \left[ a_k^E \varphi_{E,k} (\eta) e^{i \vec{k}\cdot\vec{x}} + (a^{E}_k)^\dag \varphi_{E,k}^*(\eta) e^{-i \vec{k}\cdot\vec{x}} \right]~.
\end{equation}
The creation and annihilation operators satisfy the usual properties:
\begin{equation}
a_k^E | E \rangle = 0~, \quad [a_k^E, (a^{E}_{k'})^\dag] = \delta_{k k'}~.
\end{equation}
As usual we can build a host of other states by acting with $a_k^E$ and $(a^{E}_{k})^\dag$ on $| E \rangle$. We will discuss several features of $|E\rangle$ in the following subsections.

One need not restrict themselves to the Bunch-Davies vacuum. Indeed, the `vacuum' state of the quantum field could be rather different. For instance, one could consider the $| \text{out} \rangle$/$| \text{in} \rangle$ states which are annihilated by positive frequency modes in the far future/past, i.e. modes which behave as $\sim (-\eta)^{1/2 - \nu}$ as we approach $\mathcal{I}^+$/$\mathcal{I}^-$, or a more general family known as the $\alpha$-vacua \cite{Mottola:1984ar,Allen:1985ux}, parametrized by a complex number $\alpha$. These vacua and their properties are reviewed in \cite{Bousso:2001mw,Spradlin:2001pw}.

\subsection{Two-point functions}\label{twopointfunctions}

One can also define the Wightman $G_W(x,x')$ function for the field $\Phi(x)$. This solves the equation:
\begin{equation}\label{wightman}
\nabla^\mu \nabla_\mu G_W(x,x') - m^2 G_W(x,x') = 0~.
\end{equation}
If we demand that $G_W$ be de Sitter invariant, it follows that it can only depend on the de Sitter invariant distance between two points $d(x,x') = \ell \arccos P(x,x')$. For planar coordinates one finds $P(x,x') = (\eta^2 + \eta'^2-(x-x')^2)/2(\eta\eta')$. Thus we obtain two de Sitter invariant solutions to (\ref{wightman}) which are given by hypergeometric functions. In terms of $P$, $G_W$ obeys:
\begin{equation}\label{green}
(P^2-1) \frac{d^2 G_W}{dP^2} + (d+1) P \frac{d G_W}{d P} + m^2 \ell^2 G_W = 0~,
\end{equation}
where $(d+1)$ is the dimensionality of spacetime. 

In addition to de Sitter invariance, we must specify one additional condition to fully fix a solution. This condition can be related to the presence of singularities of $G_W$ as a function of $P$. All solutions to (\ref{green}) have singularities at the antipodal point $P=-1$ except one. The one with no antipodal singularity is the Green function arising from Bunch-Davies vacuum state $|E\rangle$.\footnote{These are also the Green functions that arise from analytic continuation of the Green function on the $(d+1)$-sphere, which is Euclidean de Sitter space. See \cite{Higuchi:2010xt} for a discussion.} In terms of the Euclidean modes $\Phi^E(x) = \eta \varphi^E(x)$ in (\ref{Emodes}) we can write the Euclidean Green function $G^E_W$ as:
\begin{equation}\label{sandwich}
G^E_W (x,x') = \langle E | \hat{\Phi}^E (x) \hat{\Phi}^E (x') | E \rangle = \frac{\Gamma[h_+]\Gamma[h_-]}{(4\pi)^{(d+1)/2} \Gamma[(d+1)/2]}F\left( h_+,h_- \frac{d+1}{2} ; \frac{1+P(x,x')}{2} \right)~,
\end{equation}
where $h_\pm \equiv d/2 \pm i\sqrt{m^2 \ell^2 - d^2/4}$. The short distance singularities of $G^E_W(x,x')$ reproduce those in flat space and as was mentioned, it contains no antipodal singularities.

One may also wish to consider the de Sitter invariant two-point function which solves (\ref{green}) but behaves purely as $\sim (-\eta)^{1/2 + \nu}$ or $\sim (-\eta)^{1/2 - \nu}$ near $\mathcal{I}^+$. For $\nu \in \mathbb{R}$, such a two-point function is not given by an expression such as (\ref{sandwich}) for any normalizable state and is referred to as a two-point function with {\it future boundary conditions} \cite{Anninos:2011jp,Ng:2012xp}. This two-point function, unlike the two-point functions arising from any of the normalizable $\alpha$-vacua, is related by an analytic continuation to the two-point function in the canonical vacuum (with respect to the global timelike Killing vector) of AdS$_4$ \cite{Anninos:2011jp}. As we will discuss in section \ref{fullyquantum}, these are the two-point functions related to the dS/CFT correspondence. Notice that the future boundary condition two-point functions have singularities at the antipodal point.

\subsection{Retarded Green function}

\begin{figure}[h!]
{\begin{center}
   \includegraphics[height=70mm]{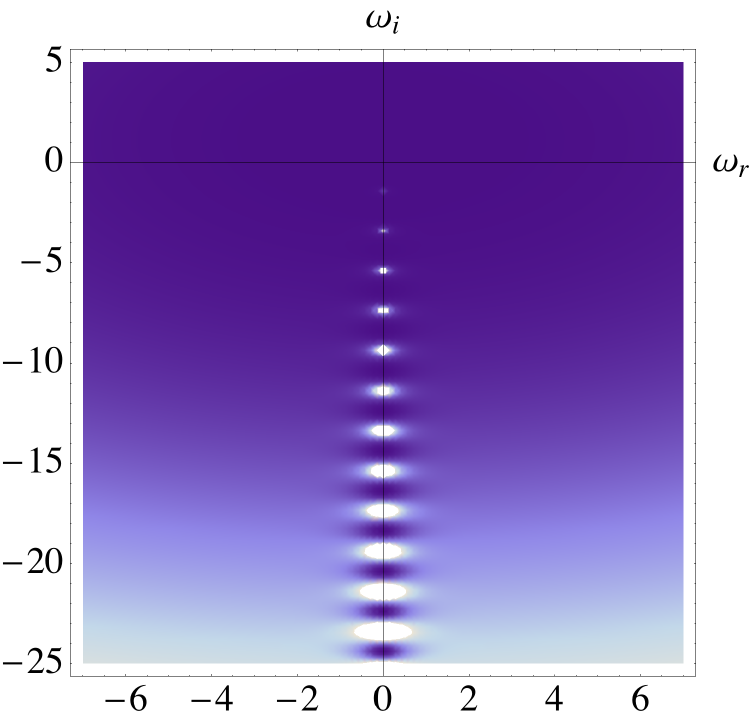}
   \end{center} 
\caption{Density plot of absolute value of $\tilde{G}_R(\omega)$ in the complex $\omega$-plane for $m^2 \ell^2 = 1$, $l = 1$.}
\label{fig:qnmdsfigs}
  }
\end{figure}

We can also compute retarded correlators, particularly near the worldline of the observer \cite{Anninos:2011af}, i.e. $r=0$ in static patch coordinates. These obey an equation:
\begin{equation}
\left( \nabla^\mu \nabla_\mu  - m^2 \right)G_{R}(t;(r,\Omega),({r',\Omega'})) = \frac{1}{\sqrt{-g}}\delta(t)\delta(r-r')\delta(\Omega-\Omega')~,
\end{equation}
with $G_{R} = 0$ for $t < 0$. It is convenient to perform the coordinate transformation $r = \ell\tanh y$. Consider a scalar field $\Phi(t,r,\Omega) = (\tanh y)^{-1} \chi(t,y,\Omega)$ with initial profile of $\chi(0,y,\Omega)$. We can express its time evolution as:
\begin{equation}
\chi(t,x) = \int d x' \; G_{R}(t;x,{x'}) \; \partial_t \chi(0,x') +  \int d x' \; \partial_t G_{R} (t;x,{x'}) \chi(0,x')~, \quad x \equiv (y,\Omega)~.
\end{equation}
We must specify conditions for the $\chi(t,x)$. In direct analogy with the case of black holes \cite{Ching:1994bd}, we consider solutions to the wave equation with no incoming flux from the past cosmological horizon, which we call $\chi_{out}(t,x)$, and solutions which are regular near the origin, which we call $\chi_{reg}(t,x)$. Then one can show that the Fourier transform in time of $G_R$ (expanded in spherical harmonics $Y^{lm}(\Omega)$) is:
\begin{equation}
\tilde{G}^{lm}_R (\omega,y,y') = \frac{\tilde{\chi}^{lm}_{reg}(\omega,y)\tilde{\chi}^{lm}_{out}(\omega,y')}{W(\omega)}~, \quad y < y'~,
\end{equation}
and with $\tilde{\chi}^{lm}_{reg}$ and $\tilde{\chi}^{lm}_{out}$ switched for $y > y'$. The function $W(\omega,l,m) = \chi_{reg}^{lm} \partial_y \chi^{lm}_{out} - \chi^{lm}_{out} \partial_y \chi^{lm}_{reg}$, known as the Wronskian, is independent of $y$. The zeros of $W(\omega,l,m)$ are the poles (in the complex $\omega$ plane) of $G_{R}$ and are nothing more than the quasinormal modes discussed in section \ref{qnmsec}, see figure \ref{fig:qnmdsfigs} for an example. In \cite{Anninos:2011af} the structure of $G_R$ near the worldline where $y,y' \to 0$ was studied. It was found that the `worldline correlators' exhibit a hidden $SL(2,\mathbb{R})$ symmetry. Though not an isometry of de Sitter space itself, this hidden $SL(2,\mathbb{R})$, reminiscent of similar hidden symmetries found upon studying the wave-equation in a Kerr background \cite{Castro:2010fd}, is related to particular conformal symmetries of dS$_4$. A particularly simple example is that of a conformally coupled scalar with $m^2 \ell^2 = 2$ for which:
\begin{equation}
\lim_{r,r' \to 0} {G}^{lm}_R(t) \propto \theta(t) \left( {\sinh \frac{t}{2\ell}} \right)^{-(2l+2)}~,
\end{equation}
where we recall $l$ is the total angular momentum on the $S^2$. This is precisely the form of the retarded Green function of a $(0+1)$-dimensional $SL(2,\mathbb{R})$ invariant theory, i.e. conformal quantum mechanics \cite{de Alfaro:1976je}. The $SL(2,\mathbb{R})$ symmetry arises from the fact that dS$_4$ is conformally equivalent to AdS$_2 \times S^2$ -- it is the isometry group of the AdS$_2$.

Of course the boundary conditions we have imposed on $\chi$ are by no means unique and we may consider allowing radiation from the past horizon if we wish. Furthermore, we may consider defining more general conditions at some surface $r = \epsilon$ away from the worldline. In such a way one could construct a more general class of worldline correlators. 

\subsection{Wavefunctionals}\label{wavefunctionals}

A complementary approach to constructing quantum states is to build a transition amplitude between some initial and final configuration of the field $\Phi(x)$. For instance, we can consider the path integral of a field which is a positive frequency mode with respect to the Bunch-Davies vacuum at some very early time and ends in some late time configuration $\phi(\vec{x})$. The condition that the modes are born out of $| E\rangle$ amounts to a `no boundary condition' in the far past where $k |\eta| \gg 1$, which can be achieved by adding a small imaginary part to $\eta$. The path integral at some late time $|\eta_0| \ll 1$ becomes:
\begin{equation}\label{wavefunction}
\Psi_E[\eta_0;\phi] \propto e^{i S[\eta_0;\phi]}~,
\end{equation}
where $S[\eta_0;\phi]$ is the late time part of the action evaluated on-shell and the constant of proportionality does not depend on $\phi$. The absolute value squared of the wavefunctional may be interpreted as the probability $\mathcal{P}[\phi]$ for a given late time configuration $\phi$. 

Let us consider the case of a massless scalar field in planar coordinates. Modes satisfying our boundary conditions and solving (\ref{eom}) are given by:
\begin{equation}
\Phi (\eta,\vec{x}) = \int \frac{d^3 {\vec{k}}}{(2\pi)^3} \; e^{i \vec{k} \cdot \vec{x}} \phi_{\vec{k}} \frac{(1 - i k \eta)e^{i k \eta} }{(1 - i k \eta_0) e^{i k \eta_0}}~.
\end{equation}
We can thus compute the probability distribution coming from the wavefunction (\ref{wavefunction}): 
\begin{equation}
\mathcal{P}[\phi] \propto e^{-2 \int d^3{\vec{k}} \; \beta_k |\phi_{\vec{k}}|^2}~, \quad \beta_k = \frac{\ell^2 k^3}{2(1+k^2\eta_0^2)(2\pi)^3}~.
\end{equation}
Using the above wavefunctional, one can for example compute the late time Wightman function of a massless field in the Bunch-Davies vacuum by performing the integral: 
\begin{equation}
\lim_{\eta,\eta' \to 0} \langle E | \Phi(\eta,\vec{x}) \Phi(\eta',\vec{y}) | E \rangle = \int \mathcal{D}\phi |\Psi_E [0;\phi]|^2 \phi(\vec{x})\phi(\vec{y}) = \frac{1}{2\ell^2} \int \frac{d^3 \vec{k}}{(2\pi)^3} \; {e^{i \vec{k} \cdot (\vec{x}-\vec{y})}}\frac{1}{k^3}~. 
\end{equation}
Notice that to achieve the above correlator we had to integrate over the late time configurations. In the same way, one can (perturbatively) construct wavefunctionals for metric perturbations and other matter fields about a fixed de Sitter background. 

One may also consider the two-point functions obtained by taking variational derivatives of $\Psi_E[\phi]$ with respect to the boundary values of the fields. Focusing again on the case of the massless field, it is a straightforward calculation to show that \cite{Maldacena:2002vr}:
\begin{equation}
\frac{\delta^2 \Psi_E [0;\phi_{\vec{k}}]}{\delta \phi_{\vec{k}_1} \delta \phi_{\vec{k}_2} }  = - (2\pi)^3 \ell^2 k_1^3 \delta \left(\vec{k}_1 + \vec{k}_2 \right)~.
\end{equation}
These two-point functions are nothing more than the future boundary two-point functions discussed in the previous subsection. By taking $n$ variational derivatives of $\Psi_E$, one can in fact view  $\Psi_E$ as the generating functional of $n$-point functions with future boundary conditions.

\subsection{Branching Diffusion}\label{branching}

We would like to make one final comment about the evolution massless fields. Consider an initial configuration of a massless field in a fixed de Sitter background. As time passes, space expands and eventually some of the modes of the quantum fluctuations about  the classical field configuration grow large compared to the horizon scale. These modes freeze, classicalize and no longer communicate with the sub-horizon modes. The quantum fluctuations of the frozen modes in turn grow large as the space further expands and eventually themselves freeze and classicalize. This branching diffusion process \cite{Starobinsky:1982ee} continues all the way up to $\mathcal{I}^+$, thus populating it in one of the plethora of possible late configurations. In fact the correlation functions for massless fields in the Bunch-Davies vacuum grow logarithmically for large distances and thus there is no notion of cluster decomposition for these fields in a de Sitter background. Such an effect, in contrast, does not occur in AdS$_4$ or four-dimensional flat space. We will return to the question of how this `landscape' of late time configurations is organized in section \ref{trees}.

\section{de Sitter Space Semiclassical}

Having discussed some classical and (slightly) quantum results, we proceed to give a brief account of several semiclassical aspects. These are quantum phenomena that cannot be accounted for from a perturbative analysis.

\subsection{Thermodynamics}

The laws of black hole thermodynamics \cite{Bardeen:1973gs,Bekenstein:1973ur} were a crucial stepping stone paving the way toward a more complete understanding of the microscopic nature of a black hole. The basic observation was that classically physical processes can only lead to an increase in the area $A_{BH}$ of the black hole horizon (area law) and the way the area responds to classical processes follows an equation equivalent in form to the first law of thermodynamics, namely:
\begin{equation}
\delta M = T_{BH} \delta S_{BH} + \Omega_H \delta J~,
\end{equation}
where $\delta M$ is the change in the ADM mass, $T_{BH}$ is the surface gravity divided by $2\pi$, $\Omega_H$ is the angular velocity of the horizon, $\delta J$ is the change in ADM angular momentum and $S_{BH}  = A_{BH}/4G$. Hawking famously discovered that $T_{BH}$ is in fact the temperature of the black hole as measured by a far away observer. The appearance of the first law was no coincidence and a statistical mechanics interpretation was eventually provided by string theory \cite{Strominger:1996sh}. The black hole entropy is interpreted as a count of the number of microstates with macroscopic quantities equivalent to those of the black hole they constitute.

In analogy to the case of black holes, it was proposed by Gibbons and Hawking \cite{Gibbons:1977mu} that there exists an entropy associated to the pure de Sitter horizon:
\begin{equation}
S_{dS} = \frac{\pi\ell^2}{G}~.
\end{equation} 
This de Sitter entropy is proportional to the area of the horizon, as in the case of black hole entropy. Furthermore, as with all horizons, Hawking's famous result tells us that the de Sitter horizon has a temperature associated to it, given by $T_{dS} = 1/2\pi\ell$. A quick way to see this is to Euclideanize the static patch geometry by taking $t \to i t_E$ and noting that one can only avoid conical singularities in the Euclidean geometry by taking $t_E \sim t_E + 2\pi \ell$. The temperature of a system whose physical quantities are periodic in Euclidean time is given by the inverse of the periodicity. For the cosmological constant measured in our own universe one finds $S_{dS} \sim 10^{120}$, which far exceeds the present entropy\footnote{The entropy of the present day universe, including the Bekenstein-Hawking entropy of supermassive black holes, is estimated around $10^{104}$ (see \cite{Egan:2009yy} and references therein).} of the remaining matter content.

When studying black hole thermodynamics, one typically asks how the black hole responds to some classical process such as absorbing some mass. What is the analogous question in the case of a cosmological horizon surrounding the static patch observer? We begin with some mass $M$ localized at the center of the static patch. As can be seen from (\ref{schw}), this has the effect of {\it reducing} the size of the cosmological horizon. One can indeed check that if this mass falls outside the cosmological horizon:
\begin{equation}
- \delta M = T_{dS} \delta S_{cos}~.
\end{equation}
This is the analogue law of thermodynamics for a de Sitter horizon. It can be naturally generalized to include angular momentum and other conserved quantities. The point is that the de Sitter horizon responds to physical processes very much like any other horizon. Notice however the minus sign in front of $\delta M$. The entropy increases when we throw mass {\it outside} the cosmological horizon surrounding us. Indeed, the less information we have about the interior of the cosmological horizon, the higher its entropy will be.  

We end with an important remark. Though it seems that much of the black hole picture carries forward to the case of the de Sitter horizon, there are some crucial differences. For instance, an observer can never approach her own cosmological horizon and probe it. In other words, there is no sense in which the cosmological horizon is an object localized in space as would be the case for a black hole. Also, black holes in flat space decay when emitting Hawking radiation. On the other hand, at least naively, the Hawking radiation of a de Sitter horizon is reabsorbed by the de Sitter horizon itself, leading to no overall evaporation of the horizon.

\subsection{Black Hole Nucleation}

As we discussed in section \ref{classics}, de Sitter space is classically stable. Small deformations of initial data do not have a large impact on the geometry at $\mathcal{I}^+$. Asymptotically flat space at zero temperature is also classically stable in Einstein gravity. On the other hand, the quantum mechanical story may be quite different. For instance observers in the static patch are surrounded by a thermal bath and one may expect de Sitter space to mimic hot flat space which classically exhibits the Jean's instability \cite{Gross:1982cv}. Remarkably, there is no analogue of the classical Jeans instability in de Sitter space \cite{Ginsparg:1982rs}, at least perturbatively. Semiclassically however, de Sitter space is unstable toward the nucleation of Nariai black holes \cite{Ginsparg:1982rs}. One can compute the likelihood of nucleating a Nariai black hole by evaluating the on shell action of the $S^2\times S^2$ solution of Euclidean gravity with positive $\Lambda$. The $S^2\times S^2$ solution (or Euclidean Nariai space) is found to have a negative eigenvalue. It has thus been interpreted as an instanton mediating the nucleation of Nariai black holes \cite{Ginsparg:1982rs}. The nucleation rate per unit volume is found to be:
\begin{equation}
\lambda \sim  e^{-\pi/\Lambda G}~.
\end{equation}
Once the black hole is nucleated, it will decay by the emission of Hawking radiation and one estimates that in the semiclassical approximation $\Lambda G \ll 1$ the evaporation rate of the Nariai black hole far exceeds its nucleation rate \cite{Bousso:1997wi,Niemeyer:2000nq}. Thus de Sitter space is never `consumed' by Nariai black holes in the semiclassical limit. 

\subsection{Hartle-Hawking Wavefunctional}\label{hhsec}

One of the great challenges in defining a theory of quantum gravity is that (at least in four and more dimensions) there is no known controlled way of performing the path integral over geometries. On the other hand, the sum over geometries may be dominated by the saddle points of the action which we may have access to. 

Here we discuss, briefly, a proposal for the ground state wavefunctional due to Hartle and Hawking \cite{Hartle:1983ai}. For simplicity, we restrict ourselves just to the metric degree of freedom. They proposed that the groundstate wavefunctional $\Psi_{HH}[h_{ij}]$ of the Wheeler-de Witt equation, as a function of three-metrics $h_{ij}$ on some spacelike slice, is given by computing the path integral over all compact Euclidean geometries:
\begin{equation}\label{hh}
\Psi_{HH}[h_{ij}] = \int_{\mathcal{M}_{g|h}} \mathcal{D} g \; e^{-S_E[h_{ij};g]}~.
\end{equation}
These compact geometries contain a single boundary, whose real induced metric is given by $h_{ij}$. By summing over Euclidean geometries which are compact, one avoids having to specify any initial conditions and in this sense $\Psi_{HH}[h_{ij}]$ has been interpreted as giving the probability of a particular configuration $h_{ij}$ to come out of `nothing'. The proposal is inspired by the construction of the ground state wavefunction from a Euclidean Wick rotation in ordinary quantum mechanics, where the Euclidean action is assumed to vanish in the far (Euclidean) past.

The Euclidean action is:
\begin{equation}
-S_E = \frac{1}{16\pi G}  \int_{\mathcal{M}} d^4 x \sqrt{g} \left( R - 2\Lambda \right) + \frac{1}{8\pi G} \int_{\partial \mathcal{M}} d^3x \sqrt{h} K~, \quad \Lambda > 0~,
\end{equation}
where $K$ is the trace of the extrinsic curvature of the boundary three-metric with respect to an outward pointing normal vector to $\partial\mathcal{M}$. The most symmetric solutions extremizing the action are the four-sphere $S^4$ and the product $S^2 \times S^2$. We may smoothly glue these Euclidean geometries at the equator of the $S^4$ or one of the $S^2$'s onto the Lorentzian solutions dS$_4$ or dS$_2 \times S^2$ respectively at the $\tau = 0$ spatial slice of the global geometry. By `smoothly gluing' we mean that both the induced metric and extrinsic curvature should match across the boundary. Evaluating the on-shell action on these Euclidean hemispheres we find that the four-hemisphere has the smallest action and hence will be the dominant contribution to the Euclidean path integral (\ref{hh}) in the $\Lambda G \to 0$ limit. Gluing the Euclidean hemisphere $\tilde{S}^4$ onto pure global dS$_4$ (\ref{global}) at $\tau = 0$ requires $h_{ij} dx^i dx^j = d\Omega_3^2$. We can estimate semiclassically for this case that $\Psi_{HH}[\tilde{S}^4] \sim  e^{3\pi/2\Lambda G}$. On the other hand, the product of a two-hemisphere $\tilde{S}^2$ with an $S^2$ has $\Psi_{HH}[\tilde{S}^2\times S^2] \sim  e^{\pi/\Lambda G}$. One can interpret this as saying it is more likely (in the $\Lambda G \to 0$ limit) to have a pure dS$_4$ universe come out of nothing than a dS$_2\times S^2$ universe \cite{Bousso:1996au}. 

In order to get more general results one can consider minisuperspace approximations where one restricts the path integral over a limited set of geometries. For example, we might consider evaluating the path integral (\ref{hh}) over compact geometries whose boundary is a three-sphere with size $a$. A saddle point approximation of this case was achieved in \cite{Hartle:1983ai}. For $a < \ell$, the saddle that dominates is given by the lesser part of the four hemisphere with $\theta < \theta_0$ such that the three-sphere at $\theta_0$ has size $a$. Thus, the semiclassical approximation to the Hartle-Hawking wavefunction in this setup is found to be:
\begin{equation}\label{smallhh}
\Psi_{HH} [a] =  \mathcal{N} e^{\frac{\ell^2 \pi}{2 G}\left[ 1 - \left(1 - a^2 \ell^{-2} \right)^{3/2} \right]}~, \quad a \ell^{-1} < 1~,
\end{equation}
where $\mathcal{N}$ is an $a$ independent normalization factor. Notice that for $a < \ell$, $(\ref{smallhh})$ decreases exponentially with decreasing $a$, as one would expect since the size of the three-sphere in Lorentzian de Sitter is always greater than $\ell$ and hence small spheres are not classically allowed. For $a  > \ell$ there are no real contours since the size of a three-sphere in the four-sphere can be at most $a = \ell$, and instead one considers the complex extremum of smallest real action to find:
\begin{equation}
\Psi_{HH} [a] =  \mathcal{N} \cos \left[ \frac{\ell^2 \pi}{2 G} \left(a^2 \ell^{-2} - 1 \right)^{3/2} - \frac{\pi}{4} \right]~, \quad a \ell^{-1} > 1~.
\end{equation}
It is a challenging task to compute the Hartle-Hawking wavefunction in more general situations, progress in this direction includes \cite{Banks:1984np,Maldacena:2002vr,Harlow:2011ke,Hertog:2011ky} where attempts have been made to connect the Hartle-Hawking wavefunction to the partition function of the CFT dual to an auxiliary anti-de Sitter space.

\subsection{Coleman De Luccia Bubbles}\label{bubble}

One final interesting semiclassical property of de Sitter space regards the process of Coleman-de Luccia bubble nucleation \cite{Coleman:1980aw}. This occurs when the de Sitter universe at study is metastable with nearby lower minima of some scalar potential. Such a situation is observed for the known de Sitter constructions in string theory. In particular, one could study the nucleation of a de Sitter bubble inside a false de Sitter vacuum. 

In the simplest case, we can consider a scalar field $\varphi$ minimally coupled to gravity and a scalar potential $V(\varphi)$ with two positive energy minima with energies $V_+ > V_- >0$. Let us assume life starts in the false vacuum $V_+$. Then, by finding a Euclidean bounce solution $B$ which interpolates between the two Lorentzian vacua, one can estimate the decay rate for the nucleation of true vacuum bubbles in the false vacuum. The decay rate (per unit volume per unit time) is given by:
\begin{equation}
\nu \sim e^{- S_B}~,
\end{equation}
where $S_B$ is the on-shell Euclidean action of the bounce solution $\varphi_B$ with the vacuum energy density of the false vacuum subtracted. It has been noted that the ratio of decay rates between an upward transition and a downward transition between two de Sitter vacua is given by the exponential of the de Sitter entropy, which can be viewed as a manifestation of the law of detailed balance \cite{Banks:2005bm}. We should mention that the Coleman-de Luccia instanton is compact with finite Euclidean action only when the false vacuum is de Sitter space. Furthermore, when the false vacuum is flat space or anti-de Sitter space there exists a critical value of the bubble wall tension above which bubble nucleation is no longer allowed. This is never the case when the false vacuum is de Sitter space. 


\section{de Sitter Space Fully Quantum?}\label{fullyquantum}

Having discussed some classical, (slightly) quantum and semi-classical aspects of de Sitter space, we are now left with the million dollar question of whether we can propose a non-perturbative definition of de Sitter space. This is not a simple task and it is very much an open problem. The solution to the analogous problem in anti-de Sitter space follows from a beautiful conjecture \cite{Maldacena:1997re}, known as the AdS/CFT correspondence, postulating that the Hilbert space of an asymptotically anti-de Sitter universe is equivalent to the Hilbert space of a conformal field theory. Furthermore, the AdS/CFT correspondence presents a map between fields in anti-de Sitter space and the operators of the conformal field theory, as well as a prescription for computing the correlation functions of the CFT \cite{Witten:1998qj,Gubser:1998bc}. Given the striking resemblance of anti-de Sitter space and de Sitter space, one may ponder whether a similar proposal can be made. If so, it would be related to data on $\mathcal{I}^+$ which, as we discussed earlier, is inaccessible to a single observer. On the other hand, one could demand that the definition of de Sitter space involve only the data accessible to a single static patch and thus to a single observer. We will refer to these notions as {\it global holography} and {\it static patch holography} respectively.\footnote{General notions of the holographic principle for general backgrounds are elegantly discussed in \cite{Bousso:1999dw}.} It is worth pointing out that in the Lorentzian version of AdS/CFT correspondence, it is the radial direction of anti-de Sitter space that emerges holographically. In contrast, though bulk de Sitter space is also Lorentzian, it is the cosmological {\it time} direction that would emerge holographically in the case of global de Sitter holography.

\subsection{Global Holography - the dS/CFT Correspondence}\label{dscft}

In the context of global holography, there exists a natural extension of the AdS/CFT correspondence. This is known as the dS/CFT correspondence  \cite{Strominger:2001pn,Witten:2001kn,Maldacena:2002vr,Hull:1998vg}. The conjecture is that the boundary-to-boundary correlation functions at $\mathcal{I}^+$ with {\it future boundary conditions} (recall sections \ref{twopointfunctions} and \ref{wavefunctionals}) of an asymptotically $(d+1)$-dimensional de Sitter space, are in one-to-one correspondence with correlation functions of a $d$-dimensional Euclidean conformal field theory. This conformal field theory need not be unitary. Furthermore, it is proposed that bulk fields in de Sitter space are dual to single trace operators in the CFT. It is instructive to consider the case of a free scalar field $\Phi(\eta,\vec{x})$ with mass $m$ in the planar de Sitter background (\ref{planar}). At late times the scalar behaves as:
\begin{equation}
\Phi(\eta,\vec{x}) = \eta^{\Delta^+_m} \left( A(\vec{x}) + \mathcal{O}(\eta^2) \right) + \eta^{\Delta^-_m} \left( B(\vec{x}) + \mathcal{O}(\eta^2) \right)~.
\end{equation}
In direct analogy with the AdS case, it is proposed that the operator dual to $\varphi$ has conformal weight:
\begin{equation}
\Delta^\pm_m = \frac{d}{2} \pm \sqrt{\frac{d^2}{4} - m^2\ell^2}~.
\end{equation}
Whether the conformal weight of the operator is $\Delta_m^+$ or $\Delta_m^-$ depends on the future boundary condition we impose. Once we fix the boundary condition, we can read off the source and vev of the operator. For instance, if the operator has weight $\Delta_m^+$ its vev is given by $A(\vec{x})$ and its source is given by $B(\vec{x})$.

Notice that for sufficiently large $m\ell$ the $\Delta_m^\pm$ become complex, indicating that the CFT may be non-unitary (since it would contain operators with complex weights). That the CFT be non-unitary is perhaps a blessing rather than a curse. Indeed, had the CFT been unitary (or reflection positive if Euclidean) it would, according to the standard lore of AdS/CFT, be dual to anti-de Sitter space. Somehow, the non-unitarity is intricately connected to a Euclidean CFT being dual to a bulk geometry with a Lorentzian causal structure. 

Recently a concrete example of the above proposal has been conjectured \cite{Anninos:2011ui}. The bulk geometry is a theory of gravity with an infinite tower of higher integer (even) spin fields $X = \varphi, h_{\mu\nu}, w_{\mu\nu\rho\sigma}, \ldots$ These fields all interact non-linearly with each other on an equal footing \cite{Vasiliev:1990en,Vasiliev:1999ba,Iazeolla:2007wt,Vasiliev:1986td}, i.e. gravity is not much weaker than the other forces at low energies. At the level of the non-linear classical equations (which contain pure de Sitter space with all other fields switched off as a solution) no ghosts or tachyons are present in the perturbative spectrum about any solution. The parameter determining the strength of the interactions and its quantum corrections is $N \equiv \ell^2/\ell_{Pl}^2$. All higher spin fields are massless and their dual operators have a conformal weight of $\Delta_s = s+1$, where $s = 2,4,\ldots$ is their spin. The spin zero field has $m^2\ell^2 = 2$ and thus $\Delta^-_0 = 1$ or $\Delta^+_0 = 2$.

Let us choose $\Delta^-_0 = 1$ for the moment. Then we must find a theory with an infinite tower of higher even spin operators with weights $\Delta_s = s+1$, which due to the properties of higher spin fields in the bulk such as the transversality of the graviton, are also conserved. This is a rather constraining demand. The proposal, which is inspired by similar efforts for the AdS version of higher spin gravity \cite{Klebanov:2002ja,Witten,Mikhailov:2002bp,Sezgin:2002rt,hep-th/0103247,arXiv:0912.3462}, is  that the conformal field theory is simply a free theory of $N$ scalar fields $\phi^a$ transforming as an $Sp(N)$ vector. In order to match the bulk $n$-point functions one must further impose that these fields be \emph{anti-commuting} \cite{Anninos:2011ui}. Since the fields are anti-commuting scalars, they violate the spin-statistics theorem and the theory is rendered non-unitary. The tower of operators are nothing more than the higher spin conserved currents, for spin-two this is the stress tensor. The Lagrangian for the theory on $\mathbb{R}^3$ is:
\begin{equation}
\mathcal{L}_{CFT} = \int d^3 x \; \Omega_{ab} \delta^{ij} \partial_i \phi^a \partial_j \phi^b~,
\end{equation}
with a singlet constraint restricting the operator content to $Sp(N)$ singlets. Notice that the spin zero `current' $J^{(0)} = \Omega_{ab} \phi^a\phi^b$ has conformal dimension $\Delta^-_0 = 1$ and is dual to the bulk scalar. Interestingly, the $Sp(N)$ theory makes sense as a theory with a local Lagrangian description only for {\it integer} $N$, showing that the bulk cosmological constant is quantized.

One can also write down a theory for the $\Delta^+_0 = 2$ case. This is the IR CFT given by deforming the above Lagrangian by the double trace (relevant) deformation $\lambda(J^{(0)})^2$. It can be shown \cite{LeClair:2006kb,LeClair:2007iy} that the theory flows to a new CFT and a large $N$ expansion can be setup allowing one to prove in the large $N$ limit that the dimension of the spin-zero operator becomes $\Delta^+_0 = 2 + \mathcal{O}(1/N)$.

\subsection{Static Patch Holography}

The situation for a `fundamental' description of a single static patch, though crucially relevant, remains elusive. We mention here some properties that should be present in whatever this fundamental description may be. When contemplating about a fundamental description of the static patch, it is important to understand what the question we are trying to answer is. For instance, is it a theory from which we can extract the results of all possible experiments that can be performed in a single static patch? Can the formulation be sharp? This is particularly pressing due to the fact that we are unable to define precise observables within the static patch as discussed in section \ref{observables}. Recent developments related to such questions are discussed in \cite{Goheer:2002vf,Dyson:2002nt,Banks:2003cg,Parikh:2004wh,Banks:2006rx,Castro:2011xb,Alishahiha:2004md,Silverstein:2003jp,Dong:2010pm,Susskind:2011ap}. 

One property of such a theory is given by the perturbative data on the worldline, and in particular the wordline Green function. These Green functions have a particular pole structure, given by the quasinormal mode spectrum, which is known in the case of de Sitter space. Recall that a scalar field of mass $m$ has the following quasinormal mode spectrum:
\begin{equation}
\omega_n \ell = - i \left( l + 2 n + \Delta_m^\pm \right)~, \quad n = 0,1,2,\ldots
\end{equation} 
In fact, the solutions of the wave equation in the static patch are known to be hypergeometric functions and it was shown in \cite{Anninos:2011af} that a `hidden' $SL(2,\mathbb{R})$ symmetry acts on the wavefunctions.  Due to this symmetry, the worldline Green functions take precisely the form of the Green functions of a $(0+1)$-dimensional theory with $SL(2,\mathbb{R})$ invariance, such as a theory of conformal quantum mechanics \cite{de Alfaro:1976je}. This may motivate the idea that somehow the worldline of de Sitter is reminiscent of the boundary of AdS and the de Sitter horizon reminiscent to a black hole horizon in AdS. Indeed, this hidden $SL(2,\mathbb{R})$ structure is made more manifest by a conformal transformation mapping dS$_4 \times S^1$ to $\text{BTZ} \times S^2$, where by $\text{BTZ}$ we are referring to the non-rotating AdS$_3$ black hole \cite{Banados:1992wn}. The $SL(2,\mathbb{R})$'s come from the isometries of the AdS$_3$  whose quotient gives the $\text{BTZ}$.

Another property that must be reproduced is the incompressible Navier-Stokes equation obeyed by metric deformations on surfaces near the cosmological horizon. Furthermore, as noted in \cite{Susskind:2011ap}, the theory should reproduce the `scrambling' time taken for some localized distribution to spread over the entire horizon. In terms of the static patch time $t$, the scrambling time for the de Sitter horizon is given by:
\begin{equation}\label{scramble}
t_{scr} \sim \frac{1}{T_{dS}}  \log S_{dS}~.
\end{equation}
Its logarithmic dependence on the entropy may be suggestive of a matrix theory. The above equation is consistent with the exponential relaxation of the static patch as dictated by the quasinormal mode spectrum. Furthermore, the properties of black holes and Nariai black hole nucleation should be understood in the context of a fundamental description of the static patch. More particularly, one should understand the mechanism by which the de Sitter entropy is reduced when introducing mass and angular momentum inside the static patch. Significant efforts to do so, particularly in the language of a fermionic matrix model, include \cite{Banks:2003cg,Banks:2006rx,Banks:2011av}. 


Crucially, the de Sitter entropy must also be addressed. If there exists a fundamental description of the static patch, it must provide a physical meaning or precise count of the de Sitter microstates which is so far lacking. This is of particular importance given that unlike a black hole which is a localized object in space, the de Sitter horizon is observer dependent and follows the observer wherever she may go. There are also other non-perturbative issues that arise such as Poincare recurrences \cite{Goheer:2002vf,Dyson:2002nt} and the nucleation of bubbles which can be interpreted solely from the perspective of the static patch \cite{Brown:2007sd}.

\subsection{String Theory}

The search for de Sitter space in string theory is a subject in and of its own right. It is worth mentioning a few key points however. Whenever a de Sitter vacuum has been argued to exist in string theory it has been found to be {\it{metastable}} \cite{Bousso:2000xa,Silverstein:2001xn,Kachru:2003aw,Maloney:2002rr,Danielsson:2011au}. That is to say, there is a non-zero probability for that vacuum to decay to a lower energy vacuum.\footnote{This prompts the question of whether any consistent theory of quantum gravity with a de Sitter minimum must contain a landscape of vacua and whether such a de Sitter minimum is always metastable?} This can occur, for example, via bubble nucleation processes discussed in section \ref{bubble}. Thus, whatever $\mathcal{I}^+$ is in string theory, it is an object far more intricate than that of the pure de Sitter geometry. No completely stable de Sitter vacua have been constructed in string theory. On the other hand, stable Minkowski and anti-de Sitter vacua are known to exist as (supersymmetric) solutions to string theory. The ingredients required to build de Sitter vacua in string theory are not provided by supergravity on its own. Indeed, there are several no-go theorems showing that (classical) compactifications of supergravity with vanishing cosmological constant cannot contain lower dimensional de Sitter vacua \cite{Maldacena:2000mw,Douglas:2010rt}. In order to evade this, one must use ingredients in string theory such as anti D-branes, orientifolds and $\alpha'$ corrections.\footnote{See however an example of stable de Sitter vacua \cite{Fre:2002pd} in theories of gauged supergravity with non-Abelian non-compact gaugings. These theories are not known to arise from a string compactification.}

Recently, an interesting class of perturbatively stable dS$_3$ constructions was considered in \cite{Dong:2010pm}. They include classical supergravity and localized brane sources in addition to the type IIB supergravity Lagrangian. In particular, the construction includes a large number $N_1$ of D1-branes, $N_5$ of D5-branes. If these were the only ingredients, the stack of branes would have an AdS$_3 \times S^3 \times T^4$ near horizon geometry. In order to uplift the curvature and obtain a metastable positive cosmological constant for the non-compact space the authors consider adding, among other ingredients, two orientifold five-planes intersecting at a point and two sets of stringy cosmic five-branes \cite{Hellerman:2002ax} also intersecting at a point. Though near the localized brane sources curvature is strong, one may argue that the resulting net background geometry due to the backreaction of these branes can still be understood. The effect of the additional sources is to deform the original $S^3$ and the task becomes to understand the stabilization properties of the new moduli that arise due to the breaking of the $S^3$ symmetries. A few isolated solutions where all moduli are stabilized were discovered giving rise to a dS$_3$ with de Sitter length $\ell \sim 50 l_{string}$ and string coupling $g_s \sim 1/10$. It is instructive to show how the warping effects of the brane sources change the usual picture. When considering a stack of branes we can consider them to be sitting at the point of a cone with metric:
\begin{equation}
ds^2 = dw^2 + R(w)^2 ds^2_B~,
\end{equation}
In the case with only D1- and D5-branes, considered above the base space $ds^2_B$ is given by $S^3 \times T^4$ and the warp factor $R(w)$ develops a throat near the branes allowing one to decouple the AdS$_3$ near horizon. When including the additional elements that give the dS$_3$ construction, one find that the warp factor closes back up at some finite value of $w$. Such a geometry is strongly reminiscent to the de Sitter/de Sitter patch (\ref{dsds}) of dS$_3$ where $R(w) = \sin w$. In order to conserve charge one must place an equivalent (but anti) set of ingredients on the second zero of $R(w)$. These constructions will surely give further insight into the questions regarding de Sitter space.

As a final note, it is hard to believe that the difficulty/complexity of constructing parametrically large families of de Sitter vacua in string theory (especially when comparing to AdS or flat space constructions) is an accident and may be related to a far more fundamental principle. Given the generic metastability of de Sitter vacua in string theory, one is further prompted to understand the decay or exit from a de Sitter vacuum into some other FRW cosmology \cite{Sekino:2009kv,Dong:2011uf}.



\section{de Sitter Space and Beyond}

We would like to end our journey with a series of questions and ideas that remain open, elusive and exciting. As we mentioned in the introduction, de Sitter space is in many ways a tale of two observers. Hence our questions are related to both global and local confusions.

\subsection{So... what is the de Sitter entropy?}

The first question regards the nature of the de Sitter entropy. We still do not understand in what sense it is providing a count of microstates or whether it should be interpreted in a completely different fashion. It is particularly confusing due to the fact that a cosmological horizon {\it surrounds} the observer and follows her motion, in stark contrast to the case of a black hole. Should we think of the de Sitter entropy as counting the number of ways to construct a macroscopically indistinguishable static patch starting from some fundamental ultraviolet ingredients? There have been some attempts to count entropy by counting microscopic degrees of freedom in string theoretic constructions \cite{Silverstein:2003jp,Dong:2010pm} suggestive of a microstate interpretation of the entropy. It has also been interpreted as the dimensionality of the Hilbert space of quantum of de Sitter space \cite{Banks:2000fe}. Interestingly, static patch observers have fundamental limitations on how precisely they can measure physics within their cosmic horizon. Does this imply that observers themselves are mere approximations \cite{Banks:2002wr,Witten:2001kn,Harlow:2010my}? Finally we emphasize that in addition to the de Sitter entropy, one should note that the ratio of the Nariai to de Sitter entropy, i.e. the most we can reduce the empty dS entropy, is $S_{Nariai}/S_{dS} = 1/3$. This is a meaningful ratio that we should seek to understand (as well as its rotational analogue).\footnote{Curiously, it is also the ratio between the entropy of the zero temperature negative mass topological AdS$_4$ black hole and empty AdS$_4$ with $\mathcal{H}_2$ slices.}



\subsection{Holographic Projections at $\mathcal{I}^+$}

We also discussed that a possible non-perturbative definition of de Sitter space may be given by the dual CFT living at $\mathcal{I}^+$. How is the de Sitter observer reconstructed in the dual CFT at $\mathcal{I}^+$? Naively, the information that reaches $\mathcal{I}^+$ comes from many different horizons and has little to do with any single one of them. Is there a way in which we can isolate the data of a {\it single} observer to reach $\mathcal{I}^+$ and what is the meaning of this in the dual CFT? The Sachs double null problem introduced in section \ref{sachsdoublenull} may be a useful way of organizing data coming from a single static patch. We should point out that though such an isolation of data would violate global causality, such violations are not necessarily detectable by the static patch observer who is left unaltered. One motivation for trying to understand this is that a mysterious application of Cardy's formula \cite{Anninos:2009yc,Bousso:2001mw,Anninos:2011vd} has succeeded in `counting' the entropy of a single static patch. In the best of cases, this would suggest that the static patch is somehow described by a (finite entropy) thermal state in the dual CFT. Can we reproduce other aspects of the static patch such as the quasinormal modes or the cosmic near horizon fluids in the dual CFT \cite{Anninos:2011zn}?

\subsection{Cosmic Corals/Ultrametricity}\label{trees}

From a more global perspective, what are we to make of the treelike branching diffusion of de Sitter space we discussed in section \ref{branching}. Using techniques developed used to study systems with enormous configuration spaces, such as glassy systems, it was discovered  \cite{Anninos:2011kh} that the triple distance distrubution $P(d_{12},d_{13},d_{23})$ for massless fields emerging out of the Bunch-Davies vacuum sharply peaks at:
\begin{equation}\label{ultram}
d_{12} = \text{max} \{ d_{13}, d_{23} \}~,
\end{equation}
where $d_{ij}$ is the usual Euclidean distance between $\phi_i(\vec{x})$ and $\phi_j(\vec{x})$ with the zero modes removed: 
\begin{equation}
d_{ij} \equiv \frac{1}{L^d} \int d^d x \left(  \hat{\phi}_i(\vec{x}) - \hat{\phi}_j ( \vec{x}) \right)^2~, \quad \hat{\phi}_i (\vec{x}) \equiv \phi_i (\vec{x}) - \frac{1}{L^d} \int d^d x \; \phi_i (\vec{x})~, \quad x^i \sim x^i + L~.
\end{equation}
The property (\ref{ultram}), known as {\it ultrametricity}, implies a treelike organization of configurations where the leaves of the tree are the configurations and the distances are given by going up the tree. In what sense then is the bulk treelike structure \cite{Winitzki:2001np,Harlow:2011az} encoded in the ultrametricity of the boundary at $\mathcal{I}^+$? Ultrametric structures have also appeared in the study of glassy systems which are also systems with a large configuration space. A particular example is the Sherrington-Kirkpatrick model with Hamiltonian:
\begin{equation}
H_J = - \frac{1}{N} \sum^N_{i,j} J_{ij} \sigma^i \sigma^j~,
\end{equation}
where the $\sigma \in \{\pm \}$, the $i,j$ run over all $N$ lattice sites and the $J_{ij}$ are drawn from a probability distribution. Notice that there is no nearest neighbor approximation and the model is completely non-local. It was shown by Parisi that the triple overlap distribution between low temperature configurations is organized in a sharp ultrametric fashion (see \cite{spinglassbook,Denef:2011ee} for a review). Given the remarkable appearance of ultrametricity in de Sitter space, what lessons are we to draw from the physics of glasses/disordered systems? More so, what is the meaning of this ultrametric property in the dual CFT? Is this ultrametric property also present for the metric fluctuations and is it a useful way to codify the unwieldy infrared/late time behavior of de Sitter space?

\subsection{Wavefunction of the Universe}

Is it time to revisit questions about the wavefunction of the universe \cite{Hartle:1983ai,Hertog:2011ky,Maldacena:2002vr,Anninos:2011kh}? It was proposed in \cite{Maldacena:2002vr} that the late time Hartle-Hawking wavefunctional (which we discussed in section \ref{hhsec}) is computed by the partition function $Z_{CFT}$ of the dual CFT of a bulk de Sitter universe. The slow falling late time profiles of the matter fields in the bulk become sources of the dual CFT operators. Hence, full access to the wavefunctional would require an understanding of the CFT partition function with finite space-dependent sources turned on. Some of these will source irrelevant operators. Is there a controlled sense in which we can calculate this? Does $Z_{CFT}$ as a functional of its sources have the properties we would usually associate to a wavefunction, such as normalizability? A natural starting point for such questions may be the conjecture between higher spin gravity in dS$_4$ and the $Sp(N)$ theory discussed in section \ref{dscft}. It would also be of great interest to understand whether a bubble nucleation in a false de Sitter vacuum can be described purely in the language of a CFT, perhaps along the lines of \cite{Takayanagi:2011zk}.

It was observed in \cite{Maldacena:2011mk} that the Hartle-Hawking wavefunctional on the three sphere is related to the Gibbons-Hawking entropy $S_{dS}$ as:
\begin{equation}
S_{dS} = \log |\Psi_{HH}|^2 = 2 \log Z_{CFT} [S^3]~.
\end{equation}
Interestingly, the partition function of a CFT on $S^3$ has been related to the partition function of a CFT living on $\mathcal{H}_2 \times S^1$, which can be interpreted in terms of counting states of a 3d CFT on spatial $\mathcal{H}_2$ slices at finite temperature \cite{Casini:2011kv}.\footnote{This observation came up in a discussion with Tom Hartman and Daniel Jafferis.} It is unclear in what sense this counting of states is related to the de Sitter entropy. 

\subsection{Big Bang and the Emergence of Time, Bulk Unitarity} 

Interestingly, several results have been proven relating the geometry at $\mathcal{I}^+$ to the existence of singularities in the bulk spacetime \cite{Andersson:2002nr}. For instance if the topology of $\mathcal{I}^+$ is compact and the metric on $\mathcal{I}^+$ is Yamabe negative, i.e. there exists a conformal transformation making it a metric with constant negative curvature, then a singularity will exist in the past. A simple example is given by taking a compact quotient of the $\mathcal{H}_3$ in the hyperbolic slicing (\ref{h3ds}). Then one clearly encounters a big bang type singularity at $\tilde{t} = 0$ in accordance with the mathematical theorem. We may thus ask whether the partition function of the CFT dual to an asymptotically de Sitter universe makes sense on such a negative curvature compact space and to what extent this is a `resolution' of the big bang singularity. 

In addition to the big bang comes the question of time. In the AdS/CFT correspondence one observes the holographic emergence of an additional spatial dimension. On the other hand if the dS/CFT picture is correct, it would imply the holographic emergence of a cosmological time direction \cite{Strominger:2001gp} which would be a conceptual breakthrough in our understanding of holography. Cosmological evolution in time would in some way correspond to an (inverse) RG flow of the CFT with the UV fixed point being the asymptotically de Sitter phase in the infinite future. Such considerations naturally lead to the question of how bulk unitarity is encoded in the dual CFT? Other consderations of holographic cosmology include \cite{Larsen:2003pf,McFadden:2009fg}.

\subsection{Three Dimensions}

Throughout the discussion we have had very little to say about lower or higher dimensional de Sitter space. Studying lower dimensional gravity has been very useful in examining properties of quantum gravity. This is especially the case when studying AdS$_3$, which famously contains the BTZ black holes \cite{Banados:1992wn} in its spectrum. Pure three-dimensional gravity with a positive $\Lambda$ contains a dS$_3$ solution but {\it no} black holes. Instead there is a two-parameter family of conical singularities surrounded by a cosmological horizon. The size of the cosmological horizon depends on the mass and angular momentum of the conical singularity. The Euclidean version of the theory has the three-sphere and many of its quotients as smooth solutions. In \cite{Castro:2011xb} the sum over a particular class of smooth quotients of the three-sphere, known as Lens spaces, was attempted in pure gravity. It was found that the result diverges in a way that cannot be regulated using canonical field theoretic techniques. On the other hand, the same authors found that a similar computation in the context of topologically massive gravity gave rise to a finite result \cite{Castro:2011ke}. Another set of three-dimensional solutions in topologically massive gravity with a positive cosmological constant, resembling the rotating Nariai geometry (\ref{rotnariai}), were shown to contain a two parameter family of black holes \cite{Anninos:2009jt}. These may serve as useful toy models to understand properties of their higher dimensional counterparts. 

\subsection{What are the `correct' metaobservables?}

To end we would like to mention an important conceptual question that arises when considering the metaobserver. Cosmological observations are given by considering snapshots of the sky which contain data on some past spacelike slice. A particular snapshot contains the cosmic microwave background radiation spectrum of fluctuations. We typically assume that the scale invariant nature of these fluctuations is due to the presence of an early inflationary era endowed with a particular quantum state, such as the Bunch-Davies state. The initial quantum state provides us with a probability distribution of late time configurations but it does {\it not} provide us with a way to relate the physical measurements a metaobserver makes with the variables of inflationary cosmology, such as the inflaton and the metric. This is due to the the fact that the reheating surface at the end of inflation may be infinitely large (due, for example, to quantum fluctuations of the inflaton pushing it back up the potential \cite{Vilenkin:1983xq,Linde:1986fd}) and must be regulated \cite{Salem:2012ve}.

\section*{Acknowledgements}

It is a great pleasure to thank Tarek Anous, Carmen Avram, Tom Banks, Frederik Denef, Daniel Harlow, Tom Hartman, Sean Hartnoll, Daniel Jafferis, George Konstantinidis, Tongyan Lin, Gim Seng Ng, Mike Salem, Edgar Shaghoulian, Steve Shenker, Eva Silverstein, Douglas Stanford, Andy Strominger, Lenny Susskind and Gonzalo Torroba for many useful discussions. The author would also like to thank the Michigan Center for Theoretical Physics and the Kavli Institute for Theoretical Physics, Santa Barbara, where part of this work was completed. This work has been partially funded by DOE grant DE-FG02-91ER40654.

\end{document}